\documentclass[%
 reprint,
 amsmath,amssymb,
 aps,
]{revtex4-1}

\usepackage{graphicx}% Include figure files
\graphicspath{ {figs/} }
\usepackage{dcolumn}% Align table columns on decimal point
\usepackage{bm}% bold math
\usepackage{color,soul}

\DeclareMathSymbol{\shortminus}{\mathbin}{AMSa}{"39}

\begin{document}

\preprint{APS/123-QED}

\title{Enhanced repulsively bound atom pairs in topological optical lattice ladders}% Force line breaks with \\

\author{Stuart Flannigan}
%\email{stuart.flannigan@strath.ac.uk}

\author{Andrew J. Daley}
\affiliation{Department of Physics and SUPA, The University of Strathclyde, John Anderson Building, 107 Rottenrow East, Glasgow G4 0NG, UK}

\date{\today}% It is always \today, today,

\begin{abstract}
There is a growing interest in using cold-atom systems to explore the effects of strong interactions in topological band structures. Here we investigate interacting bosons in a Cruetz ladder, which is characterised by topological flat energy bands where it has been proposed that interactions can lead to the formation of bound atomic pairs giving rise to pair superfluidity. 
By investigating realistic experimental implementations, we understand how the lattice topology enhances the properties of bound pairs giving rise to relatively large effective pair-tunnelling in these systems which can lead to robust pair superfluidity, and we find lattice supersolid phases involving only pairs. 
We identify schemes for preparation of these phases via time-dependent parameter variation and look at ways to detect and characterise these systems in a lattice.
This work provides a starting point for investigating the interplay between the effects of topology, interactions and pairing in more general systems, with potential future connections to quantum simulation of topological materials.
\end{abstract}

\pacs{Valid PACS appear here}% PACS, the Physics and Astronomy
\maketitle

\textit{Introduction.~} 
Recent experiments have demonstrated the utility of ultra-cold atoms in optical lattices to explore the physics of topological quantum systems~\cite{Lin:2011aa,Struck996,PhysRevLett.108.225304,Jotzu:2014aa,PhysRevLett.107.255301,PhysRevLett.111.185302,Flaschner:2016aa,PhysRevLett.118.240403,Aidelsburger:2018aa,PhysRevLett.111.185301,PhysRevB.98.115124,Quelle:2017aa}. These systems have band structures characterised by a non-local order parameter resulting in novel global features that are in a separate classification from conventional phases~\cite{RevModPhys.88.021004,RevModPhys.82.3045}. 
While single-particle properties are generally well understood and have recently been measured experimentally~\cite{Aidelsburger:2015aa,Atala:2013aa,Asteria:2019aa,Tarnowski:2019aa,Wimmer:2017aa}, there are still many open questions relating to interacting quantum systems in these band structures, questions that cold-atom systems are perfect for exploring~\cite{RevModPhys.91.015005,PhysRevLett.110.185301,PhysRevLett.112.156801,PhysRevB.91.245135,Aidelsburger:2018ab,PhysRevB.97.201115,PhysRevLett.119.180402,PhysRevA.93.033605,PhysRevB.92.115120,PhysRevA.89.033632,PhysRevLett.117.163001}.
In this work, we investigate interacting bosons in a topological band structure where the single-particle kinetic energy is completely frustrated~\cite{doi:10.1080/23746149.2018.1473052,Liu_2014,PhysRevLett.81.5888,Mukherjee:15,PhysRevA.88.063613,PhysRevB.88.220510,PhysRevLett.108.045306,PhysRevB.82.184502,PhysRevB.97.214503,PhysRevB.98.134513,MSmith,PTorma,PhysRevA.99.023613,PhysRevA.99.023612}, and find that the topology enhances the formation of bound pairs allowing them to remain stable for higher temperatures. Going beyond the regime of perturbative interactions we find that pair superfluid phases can be engineered, prepared and detected in current optical-lattice experiments. 
This opens up ways of exploring the complex interplay between topological band structures and strongly interacting systems allowing for investigations into the effects on the many-body phases and on the resulting dynamical properties. 

% Topological bound states in the SSH model~\cite{PhysRevA.94.062704,Marques_2018} and in spin systems~\cite{PhysRevA.101.023620,PhysRevA.95.063630,PhysRevB.96.195134,Qin:2018aa}

\begin{figure}
\includegraphics[width=8.6cm]{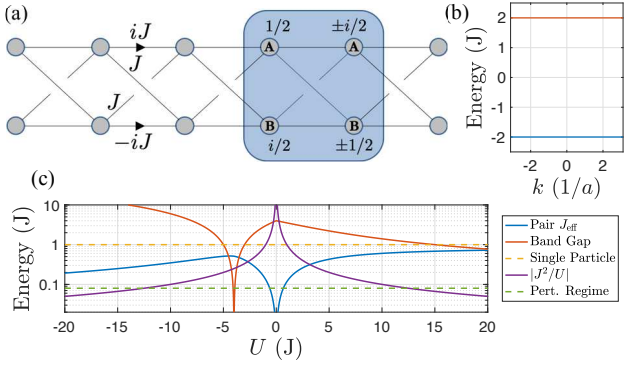}
\centering
\caption{Creutz ladder geometry (a) and band structure (b), where $a$ is the lattice spacing between the unit cells and we have highlighted in the blue box all non-zero components of the Wannier functions associated with each (highest $+$, lowest $-$) flat band. We have highlighted the two sites of the unit cell as $A$ and $B$. (c) For the case $U\equiv U_A = U_B$ a comparison of the effective nearest-neighbour pair-tunnelling in the lowest energy dispersive two-atom bound state, $J_{\rm eff}$ (blue) to that for conventional doublons with a kinetic energy, $|J^2/U|$ (purple). The green dashed line indicates an upper energy bound for the weakly interacting regime and the red line is the energy band gap between the lowest dispersive two-atom band and the first excited one, see blue and red curves in Fig.~\ref{BS_Disp}.}
\label{C_Lad}
\end{figure}

Specifically, we analyse the properties of bosons in a Creutz ladder (shown in Fig.~\ref{C_Lad}), which is characterised by complex tunnelling amplitudes along the legs of the ladder while also having diagonal tunnelling components between the legs~\cite{PhysRevX.7.031057}. In this system geometrical frustration results from the combination of these tunnelling terms where there is a destructive interference effect that completely suppresses the single-particle kinetic energy and gives rise to flat energy bands. However, it has been previously shown that including an onsite interaction can lead to the formation of bound pairs that are stable even for infinitesimal interaction strength which now have dispersion completely dictated by the interactions~\cite{PhysRevA.88.063613,PhysRevB.88.220510,PhysRevLett.108.045306,PhysRevB.82.184502,PhysRevB.97.214503,PhysRevB.98.134513,PhysRevB.98.220511,Mishra_2016,PhysRevB.98.094513,PhysRevA.100.043829,PhysRevB.101.184112,GPel,Kuno:2020aa,PhysRevResearch.2.013348,PhysRevB.98.155142,Zurita:2020aa,PhysRevB.98.184508}. There is growing interest in repulsively interacting bound pairs in general cold-atom systems~\cite{Winkler:2006aa,PhysRevA.86.013618} as well as in more novel band structures~\cite{PhysRevA.101.023620,PhysRevA.95.063630,PhysRevB.96.195134,Qin:2018aa,PhysRevA.94.062704,Marques_2018}, but pairs are usually only stable for large interaction strengths compared to the tunnelling. 
By analysing the dispersion relation for single bound pairs in the Creutz ladder beyond the limit of weak interactions we find that in contrast to those formed in conventional lattices, the pair kinetic energy grows with increasing interaction strength. This shows that the topology of the lattice plays an important role and enhances the formation and properties of these pairs.%, indicating that they are \emph{topologically enhanced bound pairs}.}

Previous ground state analysis of these systems has identified many-body phases where the correlations are dominated by superfluidity of these pairs~\cite{PhysRevA.88.063613,PhysRevB.88.220510}. Here we study the excitation spectrum and investigate the robustness of pair correlations to temperature and to excitations in time-dependent preparation. This connects directly to questions of the temperature-dependence of bound fermions in flat-band geometries, which has been recently discussed in the context of topological superconductions~\cite{PhysRevB.98.134513,PhysRevLett.117.045303}. Furthermore, for high densities we identify new lattice supersolid phases~\cite{PhysRevB.92.195149,PhysRevA.91.043614,PhysRevLett.95.033003,PhysRevA.73.051601,PhysRevA.83.051606,Wang_2012,PhysRevA.79.011602,PhysRevLett.103.225301,PhysRevLett.104.165301,PhysRevLett.103.035304,PhysRevA.80.023619}, corresponding to the coexistence of a charge-density wave (CDW) and superfluidity, but where there is no single-particle superfluidity, but rather only pair superfluid correlations. 
Additionally, we offer new perspectives in the ability to prepare and detect these phases, by first proposing an experimental preparation scheme for a pair condensate using adiabatic manipulations of the optical-lattice potential~\cite{PhysRevA.88.012334,Simon:2011aa}, which can be achieved in timescales that are reachable in current experiments.
Finally we consider experimental detection through measurements of the dynamics induced after a local quench through calculations of the dynamical structure factors, where we find substantial qualitative differences between each phase indicating that they can be resolved experimentally. % We additionally find that the supersolid behaviour is also reflected in the dynamics, with a dispersion relation that resembles the helium roton spectrum~\cite{LLSS,PhysRev.94.262}, with a linear gapless branch and a roton mode branch with a local minimum close to zero gap.

\textit{Model.~} In Fig.~\ref{C_Lad}(a) we include the Creutz ladder which has a topological band structure characterised by bands with a non-trivial Zaks phase and (for these tunnelling phases) only flat energy bands, see Fig.~\ref{C_Lad}(b).  This suppression of the dynamics arise through a destructive interference effect as atoms in the Wannier states attempt to travel through the lattice. There are various ways to produce this experimentally~\cite{PhysRevA.94.023631,PhysRevX.7.031057,Kang:2020aa}, where most proposals utilise a synthetic dimension~\cite{PhysRevLett.117.220401,PhysRevLett.108.133001,PhysRevLett.112.043001,Mancini1510,Stuhl1514,PhysRevX.9.041001,Ozawa:2019aa,HPrice,PhysRevA.95.023607} where each leg of the ladder corresponds to two atomic internal states. However including interactions will then require careful tuning of the inter-component and intra-component strengths. If there are non-zero inter-component or unequal intra-component interactions then this will result in additional terms appearing, \cite{Cr_V}, complicating the simple bound state picture presented below, however if the imbalance is small, will lead to qualitatively the same features which we demonstrate explicitly below. We also propose an alternative realisation, requiring only atoms in a single internal state confined in a dimerised ladder optical potential, so is not affected by these concerns. All the tunnelling amplitudes, with the correct phase relations, are produced from only two applied fields to facilitate multiple two-photon Raman-assisted tunnelling processes (see Appendix A).

Explicitly the Hamiltonian including an onsite two-body contact interaction, $U_A$ for the $A$-sites and $U_B$ for the $B$-sites, is given by ($\hbar=1$),
\begin{equation} \label{TB_MOD}
\begin{split}
H =  \sum_n J & \left[\hat{b}_n^{\dagger} \hat{a}_{n+1}+\hat{a}_n^{\dagger} \hat{b}_{n+1}+i \hat{b}_n^{\dagger}\hat{b}_{n+1} \shortminus i\hat{a}_n^{\dagger}\hat{a}_{n+1}+h.c.\right] \\
 &+ \sum_n   \left[ \frac{U_A}{2}  \hat{a}_n^{\dagger} \hat{a}_n^{\dagger} \hat{a}_n \hat{a}_n + \frac{U_B}{2}  \hat{b}_n^{\dagger}\hat{b}_n^{\dagger} \hat{b}_n \hat{b}_n \right],
\end{split}
\end{equation}
where $\hat{a}_n^{\dagger}$  ($\hat{b}_n^{\dagger}$) creates a particle on the $A$ ($B$) site in the $n$th unit cell. In order to analyse the bound states in this system, it is advantageous to apply a basis transformation to the local Wannier basis that diagonalises the single-particle Hamiltonian. This basis is shown in blue in Fig.~\ref{C_Lad}(a) and the transformation is,
\begin{equation} \label{Basis}
\begin{split}
\hat{a}_n & = \frac{1}{2} \hat{W}_n^+ + \frac{1}{2} \hat{W}_n^- - \frac{i}{2} \hat{W}_{n+1}^+ + \frac{i}{2} \hat{W}_{n+1}^- \\
\hat{b}_n & = -\frac{i}{2} \hat{W}_n^+ - \frac{i}{2} \hat{W}_n^- + \frac{1}{2} \hat{W}_{n+1}^+ - \frac{1}{2} \hat{W}_{n+1}^-,
\end{split}
\end{equation}
where the $\hat{W}_n^{\pm}$ annihilate a boson at unit cell $n$ in the higher/lower band. This transformation allows us to explicitly see that the single-particle dynamics are suppressed,
\begin{equation}\label{Ham}
\begin{split}
H = -2J &\sum_n \hat{W}_n^{-\dagger} \hat{W}_n^- + 2J \sum_n \hat{W}_n^{+\dagger} \hat{W}_n^+ \\
+  \frac{1}{8}& \sum_n  \left[ \frac{U_A + U_B}{4} \hat{W}_n^{\dagger}\hat{W}_n^{\dagger}\hat{W}_n\hat{W}_n \right. \\
& + \frac{U_A + U_B}{4} \widetilde{W}_n^{\dagger}\widetilde{W}_n^{\dagger}\widetilde{W}_n\widetilde{W}_n \\
& + \left( U_A + U_B \right) \hat{W}_n^{\dagger}\hat{W}_n \widetilde{W}_{n+1}^{\dagger}\widetilde{W}_{n+1} \\
& - \frac{U_A + U_B}{4} \hat{W}_n^{\dagger}\hat{W}_n^{\dagger} \widetilde{W}_{n+1} \widetilde{W}_{n+1} + h.c.  \\
& -i\frac{U_A - U_B}{2} \hat{W}_n^{\dagger}\hat{W}^{\dagger}_n \hat{W}_{n}\widetilde{W}_{n+1}  + h.c \\
&\left. -i\frac{U_A - U_B}{2} \hat{W}_n^{\dagger}\widetilde{W}^{\dagger}_{n+1} \widetilde{W}_{n+1}\widetilde{W}_{n+1} + h.c. \right],
\end{split}
\end{equation}
where $\hat{W}_n = \hat{W}_n^{+} + \hat{W}_n^{-} $ and $\widetilde{W}_n = \hat{W}_n^{+} - \hat{W}_n^{-} $. 

This non-local Wannier function basis makes two novel features of this system apparent: firstly it illustrates the vanishing single-particle kinetic energy upon diagonalising the single-particle Hamiltonian, and secondly it has illuminated the existence of strong pair-tunnelling terms, $\hat{W}_n^{\dagger}\hat{W}_n^{\dagger} \widetilde{W}_{n+1} \widetilde{W}_{n+1}$, as well as nearest-neighbour interactions, $\hat{W}_n^{\dagger}\hat{W}_n \widetilde{W}_{n+1}^{\dagger}\widetilde{W}_{n+1}$, which are both proportional to the onsite interaction strength. The effect of an imbalance in the interactions, $U_A\neq U_B$, is to introduce complex single-particle density-assisted tunnelling terms.

\begin{figure}
\includegraphics[width=8.5cm]{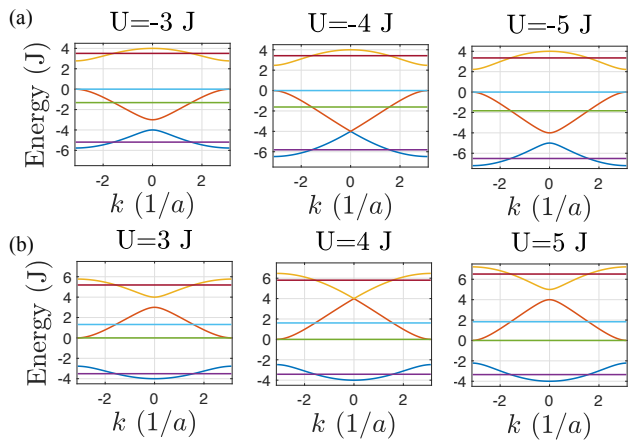}
\centering
\caption{Two-atom (single bound pair) dispersion relations for $U_A=U_B$, where each line represents the dispersion relation for the different types of two-atom bound state. Many two-atom bands now have some dispersion due to the interactions, but many others remain flat (horizontal lines). (a) Attractive interactions. (b) Repulsive interactions. Where $a$ is the lattice spacing between the unit cells. See Appendix B for the derivation of the bound state model.}
\label{BS_Disp}
\end{figure}

\textit{Enhanced bound pairs.~} If we consider the case of, $U \equiv U_A=U_B$ and only two particles, then we can apply an additional basis transformation into a two atom bound state picture. In this basis we find that the Hamiltonian can be diagonalised exactly (see Appendix B). In Fig.~\ref{C_Lad}(c) we plot the resulting dispersion of the lowest bound state band, $J_{\rm eff}$ (blue) as a function of the onsite interaction strength $U$ where there is an asymmetry between repulsive and attractive interactions, which arises through different couplings between the two single-particle bands depending on the sign of the interactions. For increasing attractive interactions the kinetic energy increases to a maximum and then begins to decrease again, agreeing with previous predictions for excitations on top of a fermion background in the Creutz ladder~\cite{PhysRevB.98.134513}. This is in contrast to the case of repulsive interactions where the kinetic energy asymptotically approaches a value very close to the single-particle tunnelling amplitude. In Fig.~\ref{BS_Disp} we plot the energy spectrum for repulsive and attractive onsite interaction strength, $U$, where changing the sign of the interactions inverts the dispersion relation meaning that the lowest energy band in the repulsive case corresponds to that of the highest band in the attractive interacting case. There are also overlaps between two qualitatively different sets of energy bands, one set dispersive, corresponding to a bound state where each single-particle is in a Wannier state centred on the same unit cell, and a set of dispersionless bands, corresponding to each single-particle centred on a nearest neighbouring unit cell. Note that for a single bound state, there are no terms that can mix these two types of state, allowing us to consider them in isolation. We will see below that in the many-body case there is mixing between these two states, the strength of which varies with density, resulting in a complex phase diagram. 

The asymmetry in Fig.~\ref{C_Lad}(c) arises from a qualitative difference in the dispersion relations for the lowest dispersive bands (blue) depending on the sign of the interactions, where for attractive interactions a Dirac cone forms in the lower bands, see Fig.~\ref{BS_Disp}(a), at the point of maximum bound state kinetic energy and then further increasing the strength of the interaction the width of this band begins to decrease due to the strong interactions mixing states within the two single-particle bands, explaining the kinetic energy observations for attractive interactions. For repulsive interactions the lowest band is not affected by this mechanism, and although there are still strong mixing between the single-particle bands, this only imposes an upper bound to the dispersion of the bound states making it possible to realise a large kinetic energy for a wide range of interaction strengths. 

In Fig.~\ref{C_Lad}(c)  we also include the effective tunnelling for pairs in a simple lattice (purple), which can be calculated in the limit that $U\gg J$ through second order perturbation theory and is $J^2/U$~\cite{PhysRevA.86.013618}. We only expect these conventional pairs to be stable in the region $|U|>10J$, and we can see for repulsive interactions in this range that the topologically bound pairs in the Creutz ladder have a kinetic energy that is larger than conventional pairs by nearly an order of magnitude. 
This difference in behaviour arises because the pairing in conventional lattices occurs with both atoms on a single site, whereas in the case of the Creutz ladder, both atoms exist in the same Wannier basis function, shown in Fig.~\ref{C_Lad}(a), allowing components to exist where each atom is bound but on different sites. These offsite binding components can then tunnel through the system due to a strong interaction induced nearest neighbour coupling between Wannier states centred at different unit cells, see Eq.~\ref{Ham}.
This has consequences for the critical temperature for superfluidity, which because it is proportional to the tunnelling amplitudes, means that the temperatures required to produce a superfluid with the pairs in this system are similar to those needed for a single-particle condensate and is an order of magnitude larger than those needed to prepare a superfluid consisting of pairs in a conventional lattice.

Even though these bound pairs exist in a novel \emph{effective} topological band structure, see Fig.~\ref{BS_Disp}, the many-body phases will be dominated by pairs condensing in the lowest bands, hence our focus on the kinetic energy of this branch. While in principle there are interesting aspects to the higher energy bands, particularly the Dirac cone at $U = \pm 4J$, we focus on the low-energy properties of the repulsively bound pairs and leave the analysis of these higher-energy features for a future study.
Nevertheless, we will see below that the enhanced properties of the lowest effective two-atom energy band leads to shorter timescales for the dynamical features, for example allowing for the preparation of states with (quasi) long range pair correlations in experimentally feasible timescales. Note that previous studies on the properties of interacting pairs in other flat band systems do not observe a strong enhancement of kinetic energy~\cite{PhysRevB.98.220511}, even in those that are also characterised by a topological invariant~\cite{GPel}. This indicates that the enhancement is not a general feature for flat band systems but is unique to the Creutz ladder geometry due to the unique form of the underlying Wannier basis, which are completely localised on only two neighbouring unit cells.

\begin{figure}
\includegraphics[width=8.5cm]{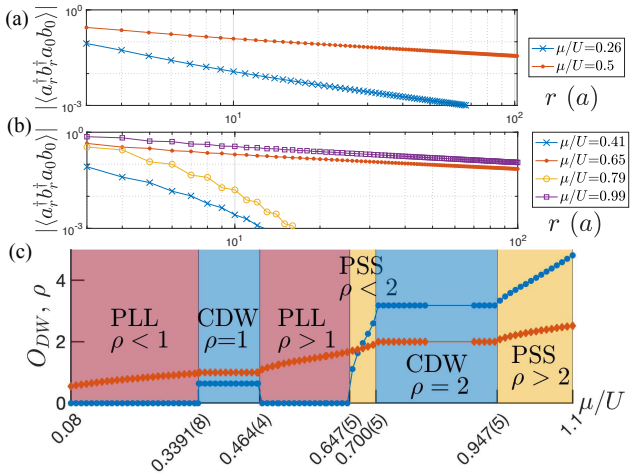}
\centering
\caption{Phases for the case $U\equiv U_A=U_B$. (a) Pair correlation functions for the pair Luttinger liquid (PLL) phases. (b) Pair correlation functions for the charge-density wave (CDW) and lattice pair supersolid (PSS) phases. (c) Phase diagram for weakly interacting bosons as the chemical potential, $\mu$, is varied relative to the interaction strength, $U$. The CDW order parameter (blue) and the density (orange) are included. The PLL, CDW and the PSS phases are indicated. Note that there is no physical significance here to the boundaries of the $\mu/U$ axis, this is simply the regime where we have focused our analysis. }
\label{Phase_Diagram}
\end{figure}

 %an infinite energy degeneracy without a well defined energy minima. 
%This results in states where all the single-particle dynamics and correlations are suppressed and means that conventional superfluidity cannot occur in these systems, leading to interesting questions regarding the properties of strongly interacting many-body phases that can be produced in these band structures.
%We find that there are now some dispersion due to the interactions breaking the energy degeneracy and there is a well defined energy minima in the lowest band allowing the system to condense into a single $k$-mode and to form into a pair superfluid. 

\textit{Phase diagram.~} We now consider the many-body bosonic case and characterise the phases that are manifested by the pairs as we vary the density. For the moment we restrict the study to interaction strengths that are symmetric $U\equiv U_A=U_B$ and that are weak compared to the separation of the single-particle energy bands, $U\ll4J$, by employing a perturbative Schrieffer-Wolff transformation~\cite{BRAVYI20112793,PhysRevA.88.063613,PhysRevB.88.220510,PhysRevB.82.184502}. This approximation allows us to focus on the lowest flat band in isolation but qualitatively preserves the main features of the full model.

We variationally calculate the ground state directly in the thermodynamic limit using a matrix product state (MPS) algorithm that assumes an infinite and uniform ansatz~\cite{PhysRevB.97.045145}, note that we increase the local dimension such that all bosonic fluctuations are captured. We include a chemical potential term, $(2-\mu) \sum_n \hat{W}_n^{-\dagger} \hat{W}_n^-$, in Eq.~\ref{Ham} and calculate the pair correlations in the site basis, $\langle \hat{a}_r^{\dagger}\hat{b}_r^{\dagger}\hat{a}_0 \hat{b}_0 \rangle$, for a range of densities controlled by the ratio of the chemical potential to the onsite interaction strength, $\mu/U$. 
In all cases the single-particle correlation function is exponentially suppressed, reflecting the lack of single-particle dispersion in this system. Additionally, we find that these off-site pair correlation functions dominate over terms of the form, $\langle \hat{a}_r^{\dagger}\hat{a}_r^{\dagger}\hat{a}_0 \hat{a}_0 \rangle$, reflecting that the pairs exist in a superposition of separate sites, in contrast to conventional repulsively bound doublons~\cite{PhysRevA.86.013618}.
The different phases are then characterised by algebraically or exponentially decaying pair correlations for the pair superfluid phase, which we refer to as a pair Luttinger liquid (PLL), and the pair CDW phase respectively, examples of which we have included in Fig.~\ref{Phase_Diagram}(a-b).
The phase diagram is shown in Fig.~\ref{Phase_Diagram}(c) where we have highlighted the PLL phases and the CDW phases and we have also plotted the value of the CDW order parameter~\cite{Rossini_2012,PhysRevB.58.R14741}, which is given by, $O_{DW} = \lim_{r\rightarrow\infty} (-1)^r \langle \delta \hat{n}_r \delta \hat{n}_0 \rangle$, where $\delta \hat{n}_i = \hat{W}_i^{-\dagger}\hat{W}_i^{-} - \rho$. Note that analysing the density-density correlation functions (with the exception of the $\rho < 1$ PLL phase, see Appendix C) the component that oscillates with wave-vector $ka=\pi$ is strongly dominant. This is due to the strong nearest neighbour interaction term in Eq.~\ref{Ham} which, as there are no longer range interactions present, ensures that this commensurate order is stabilised for all densities considered.

We agree with the predictions of Ref. \cite{PhysRevA.88.063613,PhysRevB.88.220510} which analyse the same system but restrict their analysis to low densities, where we find a PLL phase for densities between $1/2 < \rho < 1$ (per unit cell) and a phase transition to a CDW for $\rho=1$. We then investigate larger densities where we find a second PLL phase, indicating that the $\rho=1$ CDW is unstable to the addition of more pairs. And upon further increasing the density we find large regions at incommensurate density where distinct phases exist that share features of both the PLL and CDW. We denote these phases lattice pair supersolid (PSS) and are characterised by algebraically decaying pair correlations but with a non-zero density wave order parameter. 
Note that if we go beyond the weakly interacting regime such that, $U\sim J$, the coexistence of these phases is suppressed and we either have a PLL or a CDW phase with a clear phase transition point.

\begin{figure}
\includegraphics[width=8.5cm]{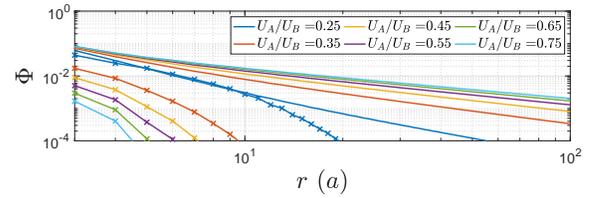}
\centering
\caption{Comparison of the pair correlations, $\Phi = |\langle \hat{a}_r^{\dagger}\hat{b}_r^{\dagger}\hat{a}_0 \hat{b}_0 \rangle|$ (solid lines) and the single-particle correlations, $\Phi = |\langle \hat{a}_r^{\dagger}\hat{a}_0\rangle|$ (crosses) for an imbalance between the onsite interactions, $U_A \neq U_B$, in the $\rho<1$ pair Luttinger liquid (PLL) phase, $\mu/U_A=0.1$.}
\label{Imb_Corr}
\end{figure}

Now we include an imbalance between the onsite interaction strengths, $U_A \neq U_B$, and perform the same analysis on the $\rho<1$ PLL phase. We plot the resulting single-particle and pair correlations in Fig.~\ref{Imb_Corr} where we see that there are still dominant algebraically decaying pair correlations, but now there are exponentially decaying single-particle correlations. For values of $U_A/U_B$ close to 1, the pair correlations greatly dominate over the single-particle correlations, indicating that the novel properties of this phase survive well into the imbalanced interaction regime. We can see that as $U_A/U_B$ decreases the single-particle correlations decay with a smaller correlation length indicating the onset of a conventional superfluid phase for $U_A/U_B \rightarrow 0$.

\textit{Experimental preparation and detection.~} Here we present a scheme to prepare a many-body eigenstate with strong pair superfluid correlations which can be achieved in a cold-atom experiment by varying the relative intensity of the lasers that create the optical potential. We begin the experimental sequence by applying a large dimerisation to the optical potential such that sites that are populated with atoms are at a much lower energy than neighbouring sites leading to atoms that are strongly localised. We then adiabatically vary the optical potential in order to slowly remove this dimerisation, which amounts to a ramp of onsite energy and allows the atoms to gradually delocalise throughout the time-dependent ramp which then prepares the eigenstate of the final Hamiltonian if the ramp time is long enough~\cite{PhysRevA.88.012334,Simon:2011aa} (see  Appendix E for details). We consider the case of large interaction strength, $U\equiv U_A=U_B=6J$, so that we have a large pair kinetic energy (see Fig.~\ref{C_Lad}(c)), allowing the correlations to spread to the entire system in timescales that are sufficiently fast so that we can ignore heating and dissipation effects. In principle, once a condensate has been prepared, we can ramp the interaction strength to weak values in order to prepare the phases predicted in the previous section. 

\begin{figure}
\includegraphics[width=8.5cm]{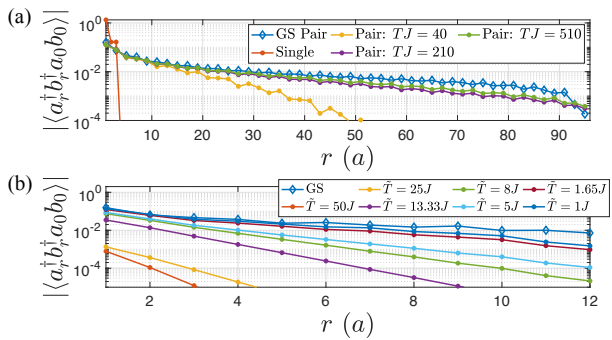}
\centering
\caption{(a) Comparison of the single-particle and pair correlations for the final produced state of the adiabatic ramp process for a system size of $M=192$ sites and $N=72$ bosons. (b) Pair correlations for finite temperatures for $M=48$ sites. For all cases, $U=6J$ and $\mu = 1.52J$.}
\label{KE_fig}
\end{figure}

In Fig.~\ref{KE_fig} we consider a ladder of $M=192$ sites and for $N=72$ bosons and plot the results of this process as the total ramp time, $TJ$, is varied. It is clear that we can produce a many-body state with significant pair correlations, $\langle \hat{a}_r^{\dagger}\hat{b}_r^{\dagger}\hat{a}_0 \hat{b}_0 \rangle$ and vanishing single-particle correlations, $\langle b_r^{\dagger} b_0 \rangle$ and $\langle a_r^{\dagger} a_0 \rangle$, in experimentally feasibly timescales ($TJ=210$). However, as we are attempting to prepare a phase that is gapless in the thermodynamic limit we expect that as we increase the system size that the total ramp time to achieve the same level of correlation decay will continue to increase. The analysis for short ramp times ($TJ=40$) indicates that for larger systems we could still produce a pair superfluid in experimentally achievable timescales with the cost of introducing effective finite size effects in the correlations.

We also consider the effects of a finite temperature on the pair correlations in the system and we use an imaginary time MPS algorithm which utilises a purification of the density matrix to calculate the state at a given temperature~\cite{PhysRevA.100.023627,PhysRevLett.93.207204}. In Fig.~\ref{KE_fig}(b) we plot the pair correlations at varying temperature and find that the correlations are exponentially suppressed at high temperatures ($\tilde{T}>20J$) where they become numerically indistinguishable from the exponentially small single-particle correlations. Note that at short distances the single-particle correlations remain qualitatively the same as those of the ground state (see. Fig.~\ref{KE_fig}(a)) where they dominate over a distance of several unit cells, simply because the particles are spread into the Wannier basis states.
For intermediate temperatures ($2J < \tilde{T}<15J$), the correlations begin to approximate those of the ground state at short distances but still decay exponentially, with a correlation length that grows as the temperature is decreased. And for temperatures $\tilde{T}<2J$ the pair correlations very closely match the zero temperature case, indicating that these pair superfluid phases are robust to finite temperatures. There are small discrepancies for the longer range tails but these will be experimentally indistinguishable.

\begin{figure}
\includegraphics[width=8.5cm]{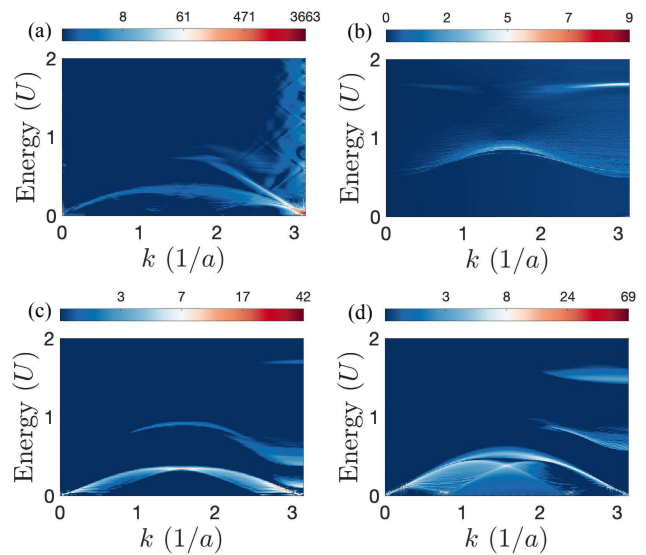}
\centering
\caption{Dynamic structure factors for the density-density correlations for each phase, for the case $U \equiv U_A=U_B$. (a) For the $\rho=1.62$, $\mu/U = 0.63$ pair Luttinger liquid (PLL) phase. (b) The $\rho=2$, $\mu/U = 0.80$ charge-density wave (CDW) phase. (c) The $\rho=1.84$, $\mu/U = 0.68$ lattice pair supersolid (PSS) phase. (d) The $\rho=2.25$, $\mu/U = 0.995$ PSS phase (d).}
\label{CDW_SS}
\end{figure}

%\textit{Experimental Detection of the Pair Supersolid.~} 
We move on and consider experimental detection of these phases. One possible way to do this is through measurements of the time-dependent onsite particle number after a local quench, in this case after the application of the number operator on a single unit cell. To this end we have calculated the dynamical structure factors for each phase (within the weakly interacting and isolated flat band limit) using an MPS algorithm for time evolving infinite systems after a local perturbation~\cite{PhysRevB.88.035103,PhysRevB.86.245107}. Explicitly we calculate the unequal time two-point correlator, $A(r,t) = \langle \Psi_0 | \delta \hat{n}_r(t) \delta \hat{n}_0(0) | \Psi_0 \rangle$, where $|\Psi_0 \rangle$ is the initial state, $\delta \hat{n}_r(t) = \hat{W}_r^{-\dagger}(t)\hat{W}_r^{-}(t) - \rho$ where $\rho$ is the density. We then take the Fourier transform which we plot in Fig.~\ref{CDW_SS} for the different phases. Note that in all cases, $U_A=U_B$.

We can see for the PLL phase (a) there is a dominant linear excitation originating from the $ka=\pi$ mode matching the predictions from Luttinger liquid theory (see  Appendix C) and for the CDW phase (b) there are well defined gapped energy branches, which can be interpreted as collective quasi-particle excitations. For the lattice supersolid phases (Fig.~\ref{CDW_SS}(c-d)) we can see gapless excitations that are linear for low energies and originate from $ka=0$ and $ka=\pi$ which is set by the underlying CDW order of the ground state, but there are also sinusoidal excitation branches present at higher energies similar to those in the CDW phase indicating the coexistence of the two phases also exists in the excitation spectrum. In Fig.~\ref{CDW_SS}(c-d) we also observe in the lower energy branches a breaking of translation symmetry as there is a doubled periodicity in $k$ space indicating that the low energy superfluid features exist on top of a dimerised ground state. Additionally in Fig.~\ref{CDW_SS}(d) there is a local minimum in the low energy dispersion which resembles the helium roton spectrum~\cite{LLSS,PhysRev.94.262,Chomaz:2018aa}, with a roton mode local minimum close to zero gap.
These calculations clearly illustrate that the distinct features of each phase are manifested in the excitation spectrum offering a way to experimentally resolve the phases through measurements of the dynamics produced after a local quench. 

\textit{Conclusion.~} We have considered the experimental opportunities of using the Creutz ladder to investigate the interplay between topological band structures and strong interactions. By analysing the properties of single repulsively bound pairs we found that the topology greatly enhances the stability and kinetic energy of formed pairs making it possible to realise and investigate pair superfluid phases in experiments with cold-atoms. We considered the ability to prepare and detect these phases where we illustrated an experimentally feasible preparation scheme allowing us to prepare a pair superfluid in realistic timescales and demonstrated that these phases can be resolved through measurements of the dynamical properties. This opens up opportunities for understanding and exploring the unique many-body phases that can be produced with strong interactions in more general topological band structures~\cite{PhysRevB.98.134513}. 

While previous analysis on other systems with flat energy bands have found similar superfluid phases dominated by bound pairs that share qualitative features as the pair superfluid phases presented here, such as in the sawtooth and Kagome lattice~\cite{PhysRevB.82.184502} or the diamond chain~\cite{PhysRevB.98.184508} they have not been considered in the context of practicalities for experimental detection. In particular, the latter system must have a very precise value for the tunnelling phases in order for it to manifest a pair superfluid, in contrast to the robustness of these phases in the Creutz ladder to similar phase deviations~\cite{PhysRevB.88.220510} and even to an interaction imbalance or finite temperatures as analysed in this paper. 
The stability of these features in the Creutz ladder is due to the enhanced properties for single repulsively bound pairs, which in contrast to the sawtooth lattice~\cite{PhysRevB.98.220511} or the diamond chain~\cite{GPel}, we find have a kinetic energy that grows with increasing interaction strength, due to the particular topology of the Creutz ladder. This distinct behaviour for the Creutz ladder opens up many questions relating to the features of other flat band systems, in particular those with additional dispersive energy bands~\cite{PhysRevLett.117.045303}.

\textit{Acknowledgements.~} This work was supported in part by the EPSRC Programme Grant DesOEQ (EP/P009565/1), and by the European Union’s Horizon 2020 research and innovation program under grant agreement No.~817482 PASQuanS. S.F. acknowledges the financial support of the Carnegie trust. Results were obtained using the ARCHIE-WeSt High Performance Computer (www.archie-west.ac.uk) based at the University of Strathclyde

All data underpinning this publication are openly available from the University of Strathclyde KnowledgeBase at https://doi.org/10.15129/81ecb9ec-7701-4940-b496-9f2785079198.

\newpage

\begin{widetext}

\appendix

% In this Supplementary Material

\section{Experimental Realisation}

Experimental realisations of the Creutz ladder have been proposed in Ref. \cite{PhysRevA.94.023631,PhysRevX.7.031057}, but require the manipulation of two internal states of an atom where each represent each leg of the ladder. In order to produce the complex tunnelling terms in these methods we need multiple Raman-assisted tunnelling processes and Floquet driving elements. The main barrier to these approaches for realising the physics analysed in this work is that the use of a synthetic dimension, with two atomic internal states, would in most cases lead to strong inter-component interactions, $U_{AB}$ and/or an asymmetry between the intra-component interactions for each species, $U_{AA}$ and $U_{BB}$. The former would lead to a strong nearest-neighbour interaction between sites in the unit cell and the latter would result in density-assisted single-particle tunnelling terms appearing between nearest-neighbour unit cells in our model. 
These additional processes would complicate the simple bound state picture that we discussed in the main text. However if an atomic species could be found that has a zero crossing for the inter-component interaction, such that $U_{AB}\approx0$ in the vicinity of a Feshbach resonance for the intra-component interaction, which results in a small interaction strength asymmetry relative to the overall magnitude, $U_{AA} - U_{BB} \ll U_{AA}$, then we explicitly illustrated that the main features of the phases survive. If these considerations can be satisfied then these schemes present viable options for realising the many-body phases proposed here. 

Here we propose a different experimental realisation with only a single internal atomic state, allowing us to easily satisfy the considerations above. Our scheme requires two fields to facilitate all Raman-assisted tunnelling processes and a two-site superlattice see Fig.~\ref{Exp_Sch}. In order to produce this, we find it convenient to redefine the tunnelling amplitudes through a gauge transformation, and we include the resulting tunnelling terms in Fig.~\ref{Exp_Sch}(a). This ladder with new tunnelling components results in the same topological physics as the ladder considered in the main text because the phases accumulated when moving around loops in the lattice are unmodified. Now the tunnelling elements along each leg of the ladder are real and have the same phase while it is the diagonal tunnelling terms that carry the complex phase factors. Notice also that there is now a dimerisation of the diagonal phase resulting in a doubled periodicity and a larger unit cell and therefore a different local Wannier function basis. The new basis is shown in Fig.~\ref{Exp_Sch}(b) and is very similar to the one used in the main text where it is also perfectly localised to two unit cells as before. This new basis results in the same single-particle spectrum and transforming the many-body Hamiltonian into this basis results in the same model that is analysed in the main text. We confirm that the Creutz ladder shown in Fig.~\ref{Exp_Sch}(a) is able to quantitatively reproduce all features presented in the main text, while also offering a more viable experimental implementation. 

For this scheme we require a particular separation of the onsite energy levels, shown in Fig.~\ref{Exp_Sch}(d) (note that there is a flexibility in the energy differences). This energy level separation can be produced with a simple $2D$ superlattice potential, however, additional potential barriers must be applied to ensure that there is no tunnelling between sites $A$-$B$ and $C$-$D$.

\begin{figure}
\includegraphics[width=17cm]{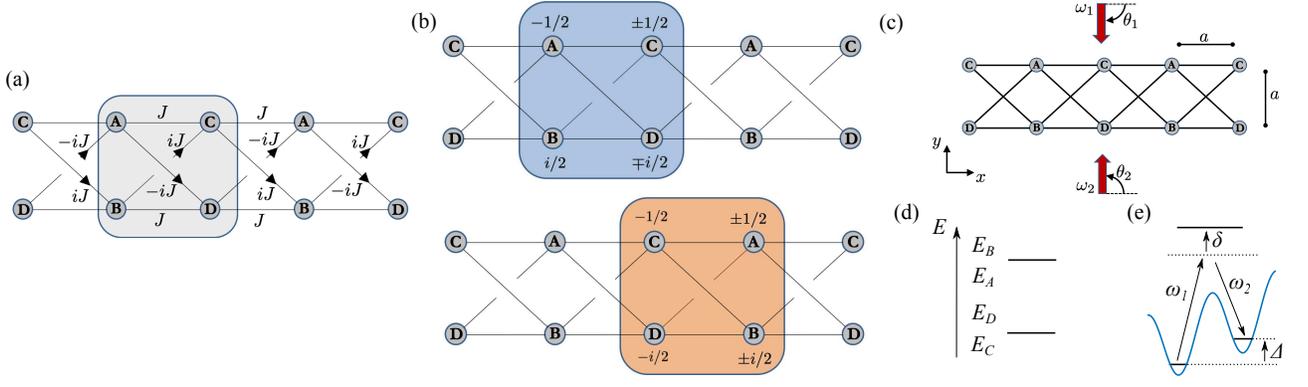}
\centering
\caption{Experimental scheme to produce the Creutz ladder in a cold-atom experiment. (a) Illustration of gauge transformed phase components for the tunnelling amplitudes, resulting in the new unit cell highlighted in grey, now consisting of four sites ($A$,$B$,$C$,$D$). (b) The non-zero components of the four new local Wannier basis functions, where for terms marked with $\pm$ the top is for the lower energy band and the bottom for the higher energy band. (c) One possible configuration for the direction of the applied fields, where $\omega_1>\omega_2$. Requiring $\lambda_1 \approx \lambda_2 = 4a$ and $\theta_1=-\pi/2$ and $\theta_2=\pi/2$ relative to the $x$-axis, see Fig.~\ref{Exp_Sch_2} for the other possible angles and wavelengths. (d) Level scheme showing the distribution of onsite energies for the lattice sites indicated in (c). (e) Illustration of the Raman-assisted tunnelling process for an optical-lattice with an energy offset between neighbouring sites.}
\label{Exp_Sch}
\end{figure}

We then apply a Raman-assisted tunnelling process, requiring two fields (Fig.~\ref{Exp_Sch}(e)), to create the two complex diagonal tunnelling processes and the real processes along the legs. This can be achieved with a single laser with a sideband allowing the necessary phase relations to be easily enforced. The effects of these applied fields is to induce tunnelling processes between off-resonant sites, \cite{PhysRevA.89.053619}
\begin{equation} \label{Gen_Eff_tunn}
J_{\alpha}=  \frac{\Omega_1\Omega_2^*}{\delta}  \int d\vec{r} \phi^*(\vec{r}) e^{i \delta \vec{k}_{\alpha} \vec{r}} \phi(\vec{r} - a\vec{R_{\alpha}}),
\end{equation}
where $\phi^*(\vec{r})$ are the onsite Wannier functions, $\delta \vec{k} = \vec{k}_1 - \vec{k}_2$ is the difference between the wave vectors of the two lasers in a single Raman process, $|\vec{k}_i| = 2\pi a/\lambda_i$, $\delta$ is the detuning between $\omega_1$, $\omega_2$ and the excited internal state (see Fig.~\ref{Exp_Sch}(e)) and $\Omega_i$ is the Rabi frequency of the applied laser with frequency $\omega_i$.

Assuming that the distance between sites, $a$, is the same in both directions, then the phase factor in the tunnelling amplitudes between each site labelled in Fig.~\ref{Exp_Sch}(c) is,
\begin{equation} \label{Eff_tunn_12}
\begin{split}
J_{CB} & \propto \exp \left( i\frac{a}{2}\left( \delta k_{x,CB} + \delta k_{y,CB} \right)\right) \\
J_{BC} & \propto \exp \left( i\frac{a}{2}\left( \delta k_{x,BC} - \delta k_{y,BC} \right)\right) \\
J_{DA} &\propto\exp \left( i\frac{a}{2}\left( \delta k_{x,DA} - \delta k_{y,DA} \right)\right) \\
J_{AD} &\propto \exp \left( i\frac{a}{2}\left( \delta k_{x,AD} + \delta k_{y,AD} \right)\right) \\
J_{DB} &\propto\exp \left(i\frac{a}{2}\delta k_{x,DB} \right) \\
J_{CA} &\propto \exp \left(i\frac{a}{2}\delta k_{x,CA} \right).
\end{split}
\end{equation}
where all terms correspond to tunnelling events from left to right in the lattice, along the directions of the arrows in Fig.~\ref{Exp_Sch}(a). The right to left processes are then the complex conjugates.

If we assume that the difference in the frequencies of the two fields in a Raman pulse is much smaller than the magnitudes, $|\omega_1 - \omega_2| \ll \omega_1,\omega_2$, where $\omega_1>\omega_2$ then we can assume that the wavelengths for each component has the same magnitude, $\lambda \equiv \lambda_{1}\approx\lambda_2$, when calculating the phases appearing in the tunnelling amplitudes. Then assuming each field is applied in a general direction in the $x$-$y$ plane, results in the phases, (where the angles are given relative to the $x$-axis)
\begin{equation} \label{phase_tunn}
\begin{split}
\phi_{CB} &= \frac{a}{2}\left( \delta k_{x,CB} + \delta k_{y,CB} \right) \approx \frac{a\pi}{\lambda}\left( \cos{\theta_2} - \cos{\theta_1} + \sin{\theta_2} - \sin{\theta_1} \right) \equiv \pi/2 \\
\phi_{BC} &= \frac{a}{2}\left( \delta k_{x,BC} - \delta k_{y,BC} \right) \approx \frac{a\pi}{\lambda}\left( \cos{\theta_1} - \cos{\theta_2} - \sin{\theta_1} + \sin{\theta_2} \right) \equiv \pi/2 \\
\phi_{DA} &=\frac{a}{2}\left( \delta k_{x,DA} - \delta k_{y,DA} \right) \approx \frac{a\pi}{\lambda}\left( \cos{\theta_2} - \cos{\theta_1} - \sin{\theta_2} + \sin{\theta_1} \right)  \equiv -\pi/2  \\
\phi_{AD} &=\frac{a}{2}\left( \delta k_{x,AD} + \delta k_{y,AD} \right) \approx \frac{a\pi}{\lambda}\left( \cos{\theta_1} - \cos{\theta_2} + \sin{\theta_1} - \sin{\theta_2} \right)  \equiv -\pi/2  \\
\phi_{DB} &=\frac{a}{2}\delta k_{x,DB} \approx \frac{a\pi}{\lambda}\left( \cos{\theta_2} - \cos{\theta_1}  \right)  \equiv 0 \\
\phi_{CA} &= \frac{a}{2}\delta k_{x,CA} \approx \frac{a\pi}{\lambda}\left( \cos{\theta_1} - \cos{\theta_2}  \right) \equiv 0,
\end{split}
\end{equation}
where on the right we have shown the values that would produce the Creutz ladder shown in Fig.~\ref{Exp_Sch}(a).
It is then simply an exercise in finding the optimal direction for the applied fields in order to produce the desired phase differences between each of the tunnelling terms. There is a huge flexibility over the relative angles and the value of the applied $\lambda$. The relationship between the angles must satisfy,
\begin{equation} \label{phase_angle}
\theta_1 = -\theta_2
\end{equation}
This requirement, means that motion in the $x$-direction does not give rise to any phase change, as $\cos{\theta_2} - \cos{\theta_1} = 0$. And so all phase dependence comes from moving in the $y$-direction with $\sin{\theta_2} - \sin{\theta_1}=-2\sin{\theta_1}$,
\begin{equation} \label{phase_tunn_y}
\begin{split}
\phi_{CB} &= -\frac{2a\pi}{\lambda}\sin{\theta_1} \equiv \pi/2 \\
\phi_{BC} &= -\frac{2a\pi}{\lambda}\sin{\theta_1} \equiv \pi/2 \\
\phi_{DA} &= \frac{2a\pi}{\lambda} \sin{\theta_1} \equiv -\pi/2  \\
\phi_{AD} &= \frac{2a\pi}{\lambda} \sin{\theta_1} \equiv -\pi/2  \\
\phi_{DB} &= 0 \\
\phi_{CA} &= 0,
\end{split}
\end{equation}
To clarify, the phases picked up from tunnelling (left to right) from site B to C, $\phi_{CB}$, and from site D to A, $\phi_{AD}$, correspond to tunnelling along the same direction (just translated in the $x$-direction). But, because B to C requires a decrease in energy while D to A requires an increase, this ensures a difference in phase between these components.  Similarly, the tunnelling (left to right) from site C to B, $\phi_{BC}$, and tunnelling from site D to A, $\phi_{AD}$, although they both require an increase in energy they involve motion in different directions, so also have a phase difference.

Now the angle $\theta_1$ and $\theta_2$ must be tuned with the wavelength, $\lambda$ in order to obtain the correct phases. As an example, we plot one possible choice in Fig.~\ref{Exp_Sch}(c) where $\theta_1=-\pi/2$ and $\theta_2=\pi/2$, which would then require a wavelength, $\lambda=4 a$. And in Fig.~\ref{Exp_Sch_2} we include all possible values for these angles and the corresponding values that $\lambda$ must be set to in order to achieve the phase values in Eq.~\ref{phase_tunn}. The values for these wavelengths are between $0$ and $4a$, so are at the same order of magnitude as the lasers responsible for producing the lattice. 

\begin{figure}
\includegraphics[width=4cm]{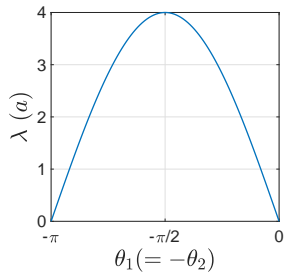}
\centering
\caption{Possible angles and wavelengths for the two fields that are responsible for producing the tunnelling processes in the dimerised Creutz ladder shown in Fig.~\ref{Exp_Sch}(c). Wavelengths given in units of the lattice spacing $a$.}
\label{Exp_Sch_2}
\end{figure}

As a final comment, if it is not possible to realise these exact phase relations (or terms with equal magnitude) for the various tunnelling terms in an experimental setting then this will most likely result in a single-particle energy band structure with bands that are not perfectly flat. If the curvature of the resulting bands are small, compared to the energy band gap and the onsite interaction strength, then some of the many-body phases illustrated in the main text have already been shown to survive into this regime~\cite{PhysRevB.88.220510}. And we believe that the new phases predicted here should also survive and that our considerations on experimental preparation and detection to still be relevant. 

% \section{Local Basis of Phase Dimerised Creutz Ladder}

\section{Bound State Model}

In this section we present the derivation of the effective single-particle Hamiltonian for interacting bound sates, which allows us to analytically calculate the single bound state dispersion relation presented in the main text. If we consider the case where the onsite interaction strengths are uniform, $U_A=U_B$, then the Hamiltonian in Eq.~\ref{Ham} does not contain any terms that correspond to the motion of a single-particle. This means that we have two types of bound state that cannot mix with one another. One set of states corresponds to two atoms on the same unit cell,
\begin{equation} \label{BS_Basis}
\begin{matrix}
\hat{\alpha}_i = \frac{1}{\sqrt{2}} \hat{W}_i^-\hat{W}_i^-, & \hat{\beta}_i = \frac{1}{\sqrt{2}} \hat{W}_i^+\hat{W}_i^+, & \hat{\gamma}_i =  \hat{W}_i^+\hat{W}_i^-, \\
\end{matrix}
\end{equation}
and the second set corresponds to the two atoms on neighbouring unit cells, 
\begin{equation} \label{BS_Basis_2}
\begin{matrix}
\tilde{\alpha}_i = \hat{W}_i^-\hat{W}_{i+1}^-, & \tilde{\gamma}_i =  \hat{W}_i^+\hat{W}_{i+1}^-\\
\tilde{\beta}_i =  \hat{W}_i^+\hat{W}_{i+1}^+, & \tilde{\kappa}_i =  \hat{W}_i^-\hat{W}_{i+1}^+. \\
\end{matrix}
\end{equation}
Transforming to this basis gives rise to a Hamiltonian containing only quadratic terms. 

If we consider the case where we only have two atoms in the system, then the basis defined by Eqs.~\ref{BS_Basis} \& \ref{BS_Basis_2} form an orthonormal basis set and we can solve the system exactly. Explicitly, the momentum space Hamiltonian is,
\begin{equation} 
H = \sum_k \Psi_k^{\dagger} H_k \Psi_k,
\end{equation}
where
\begin{equation} 
\Psi_k=\left( \begin{array}{c}
\hat{\alpha}_k\\
\hat{\beta}_k\\
\hat{\gamma}_k\\
\tilde{\alpha}_k\\
\tilde{\beta}_k\\
\tilde{\gamma}_k\\
\tilde{\kappa}_k\\
\end{array} \right).
\end{equation}
and
%\begin{widetext}
\begin{equation} 
H_k=\left( \begin{matrix}
-4 + U/4(1 - \cos k) & U/4(1 - \cos k) & iU\sqrt{2}/4\sin k & 0 & 0 & 0&0\\
U/4(1 - \cos k) & 4 + U/4(1 - \cos k) & iU\sqrt{2}/4\sin k & 0 & 0 & 0&0\\
-iU\sqrt{2}/4\sin k & -iU\sqrt{2}/4\sin k & U/2(1 + \cos k) & 0 & 0 & 0&0\\
0 & 0 & 0 & -4 + U/4 & -U/4&U/4&-U/4\\
0 & 0 & 0 & -U/4 & 4+U/4&-U/4&U/4\\
0 & 0 & 0 & U/4 & -U/4&U/4&-U/4\\
0 & 0 & 0 & -U/4 & U/4&-U/4&U/4\\
\end{matrix} \right).
\end{equation}
%\end{widetext}

In Fig.\ref{BS_Disp} we plot the energy spectrum as a function of the onsite interaction strength, $U$, where we can see that at $U=4J$ the formation and then separation of a Dirac cone, which is a signature of a topological transition. Analysing the topology of these bound states will form part of our future analysis. 

\section{Universal behaviour in the pair superfluid phases} \label{Luttinger_Liq_Map}

As we have a 1D superfluid, we expect to be able to describe the superfluid phases by mapping to a homogeneous Luttinger liquid model~\cite{Gi_Lutt}, ($\hbar=1$)
\begin{equation} \label{Lutt_equ}
H = \frac{u}{2\pi} \int dr \left[ K (\pi \Pi(r))^2 + \frac{1}{K} (\nabla \phi(r))^2 \right],
\end{equation}
where the bosonic field, $\phi(r)$, and its conjugate momentum density, $\Pi(r)$, satisfy the commutation relation, $\left[\Pi(r),\phi(r')\right]=i\delta(r-r')$. 
% $\Pi(r)$ is the conjugate momentum density of $\phi(r)$
All low energy properties of Luttinger liquids are completely known once the two parameters, $u$ and $K$, are obtained, hence the benefit of mapping our phases into this model.
In the following we extract these quantities by first fitting an algebraically decaying function to the off-diagonal pair correlation functions to obtain $K$,
\begin{equation} \label{Corr_K}
\langle W_r^{\dagger}W_r^{\dagger}W_0 W_0 \rangle \sim r^{-1/2K}.
\end{equation}
We use the pair correlation function because we know that the fundamental particles in the superfluid are pairs, and also that the single-particle off-diagonal correlation function is zero for all distances. Note that the algebraic decay persists for close to $1000$ unit cells before numerical errors destroy this behaviour, allowing us to very accurately extract the decay exponents.
We then plug the value for $K$ into the expression for the compressibility to obtain $u$,
\begin{equation} \label{u}
\kappa = \frac{d \rho}{d \mu}  = \frac{K}{u\pi},
\end{equation}
where we have evaluated the $d \rho/d \mu$ numerically from our data presented in Fig.~\ref{Phase_Diagram} from the main text.

The parameter $u$ is the effective speed of sound in the condensate which is the gradient of the linear dispersion of the excitations. The parameter $K$ controls the thermodynamic properties of the system, for example, it has been shown in Ref. \cite{PhysRevLett.68.1220} that if $K>1$ then the density transport in the system is completely robust against a single impurity, but for $K<1$ the effect of an impurity is to completely suppress transport. And in Ref. \cite{PhysRevLett.76.3192} it was shown that $K$ controls the thermal conductivity. 

There has also been recent interest in going beyond the assumptions in the Luttinger liquid theory, namely assuming that the excitations follow a linear dispersion relation. We can compute the leading order correction to this, the effective mass, $m^*$, for a non-linear Luttinger liquid theory~\cite{RevModPhys.84.1253,PhysRevA.91.043617}, through,
\begin{equation} \label{mstar}
\frac{1}{m^*} = \frac{u}{K} \frac{d }{d\mu} (u \sqrt{K}).
\end{equation}

Below, we perform this analysis for both pair superfluid phases in density ranges $\rho < 1$ and then $\rho > 1$.

\subsection{$\rho < 1$}

We begin with the $\rho<1$ superfluid phase and apply the Luttinger liquid formalism. Ref. \cite{PhysRevA.88.063613} maps this phase onto a spin $1/2$ system where $|2\rangle = |\uparrow \rangle$ and $|0\rangle = |\downarrow \rangle$, and while they test the validity of this mapping for small system sizes, we find that the mapped spin $1/2$ system does not yield the same physics as the full boson model for the infinite system considered here. We find that we need to account for the possibility of up to four bosons existing on a given site in order to properly account for all the bosonic fluctuations. Our predictions are not qualitatively different, but the critical value for the chemical potential at the phase transition is shifted. 
However, if this mapping was valid then this commensurate-incommensurate quantum phase transition at $\rho=1$ would be mathematically equivalent to that for gapped spin $1/2$ chains in a magnetic field~\cite{PhysRevB.55.5816,Gi_Lutt}. By comparing the correlation functions of the bosonic ground state to these predictions we can quantify the deviations away from the spin $1/2$ regime. If the mapping was valid, then we would be able to exactly derive the critical exponents of the phase transition on the incommensurate side simply by knowing the value of the Luttinger liquid parameters at the phase transition point. Because the deviations away from the spin $1/2$ regime are small, we can still estimate these quantities. 

We extract the Luttinger liquid parameters numerically through the procedure described above and plot these in Fig.~\ref{PLL_1}(a). The Luttinger liquid parameter $K$ is always less than one, indicating that the superfluid is dominated by charge fluctuations induced by the effective nearest-neighbour interactions, and the value at the phase transition point can be extrapolated, $K^*\approx 0.3$. 

We also plot the inverse effective mass, $1/m^*$ in Fig.~\ref{PLL_1}(a) where we see that for smaller values of the chemical potential it is much smaller than the other parameters, indicating that there are only small corrections to the conventional Luttinger liquid theory. However, for larger chemical potentials - and so larger densities - the values become negative with magnitudes that are larger than the $u$, signifying that there may be features beyond that of a Luttinger liquid.

The long distance behaviour of the density-density correlation functions for a spin $1/2$ system in the presence of an applied magnetic field is given by,
\begin{equation} \label{Corr_Dens_approx}
\langle \hat{n}_r \hat{n}_0 \rangle \sim \cos(\pi (1 + 2\delta\rho)r) r^{-2K},
\end{equation}
where $\delta\rho$ is the deviation in the spin magnetisation (or the density for the bosonic system considered here) from the commensurate regime. This correlation function predicts peaks in the Fourier transform (static structure factor) at values of $q_{\pm} = \pi(1 \pm 2\delta\rho)$. We calculate these structure factors for our system and plot these in Fig.~\ref{PLL_1}(c) for a range of $\mu/U$ and we find that the peaks do indeed correspond to the values predicted by the Luttinger liquid theory. Even though there are terms present in the model which give rise to an effective nearest neighbour interaction (see Eq.~\ref{Ham}) these are clearly negligible in this phase as the features can be predicted from the Luttinger liquid theory. This can be understood by realising that the relative magnitudes of the terms in the Hamiltonian will be dependent on density, for example $\langle W_n^{\dagger}W_n W_{n+1}^{\dagger}W_{n+1} \rangle \approx \rho^2$, and so the nearest neighbour interactions are dominated by the other processes in this phase. A similar analysis was carried out in Ref. \cite{PhysRevLett.108.045306} on a Sawtooth lattice but the commensurate phase is a $\rho=1/2$ CDW made stable by the dispersionless energy band and the incommensurate phase occurs for increasing densities. Here we find qualitatively the same features but our gapless commensurate phase is a CDW stabilised by the effective nearest-neighbour interactions. 

\begin{figure}
\includegraphics[width=17cm]{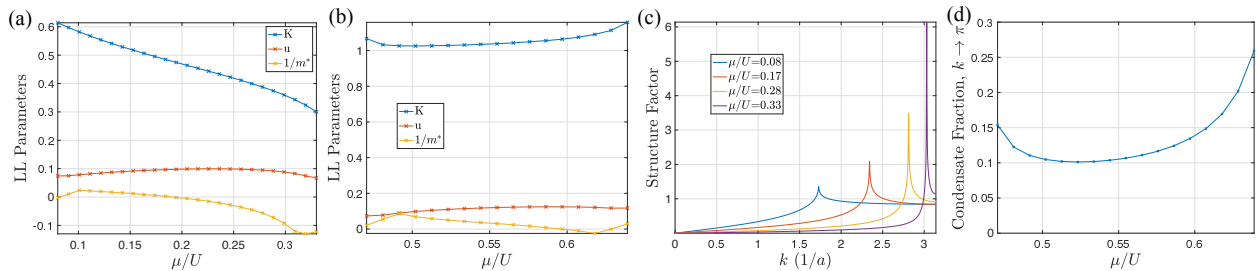}
\centering
\caption{Luttinger liquid parameters for the $\rho<1$ PLL phase (a) and for $\rho>1$ (b). (c) Static structure factors for the density-density correlation functions for $\rho<1$, on a log scale. (d) The fraction of the static structure factor in close to the $k=\pi$ mode for $\rho>1$.}
\label{PLL_1}
\end{figure}

\subsection{$\rho > 1$}

For the superfluid phase for $\rho>1$ it is not possible to map to the results from the spin $1/2$ theories, however we apply the same numerical strategy and find that the Luttinger liquid parameters are always $\geq 1$, (see Fig.~\ref{PLL_1}(b)) indicating that the phase is dominated by superfluidic fluctuations and the effects of the nearest-neighbour repulsion becomes suppressed. What is quite interesting is that there is a region around $\mu=0.5U$ where $K\approx1$ which corresponds to the case for hard-core bosons~\cite{Gi_Lutt}, indicating that we can realise a regime where the physical onsite interactions are weak but the effective onsite interactions are infinite. 

The values for the inverse effective mass approach the same magnitude as the $u$ values for smaller chemical potential values, indicating that there may be additional features present that are beyond Luttinger liquid theory. In particular we can see from Fig.~\ref{CDW_SS}(a) in the main text that the excitation spectrum in this phase is dominated by a linear branch, which is predicted from Luttinger liquid theory, but there is also clearly a sinusoidal branch present further indicating that there are additional features present here.

We also calculate the density-density structure factors and find that it is peaked at $k=\pi$ for all $\mu/U$, which is due to the strong nearest neighbour interaction in Eq.~\ref{Ham}. While in this phase these terms do not give rise to a charge density wave, they lead to features that oscillate with a given period and set the commensurate ordering in k-space. In Fig.~\ref{PLL_1}(d) we plot the fraction of the population that is in this peak, which indicates that the system is strongly condensed for larger $\mu/U$ and when $K\approx1$ the condensate peak is somewhat suppressed by the strong effective interactions.

%\textcolor{red}{As this phase manifests large corrections to the bare Luttinger liquid behaviour, as can be seen from the non-linear excitations in Fig.~\ref{CDW_SS}(a) it is not so straightforward to understand the properties through the predictions from the theory. But we can understand qualitatively why the condensate in this regime is dominated by the $ka=\pi$ mode by writing out the full expression for the density-density correlation functions~\cite{Gi_Lutt},
%\begin{equation} \label{Corr_Dens_approx}
%\langle \hat{n}_r \hat{n}_0 \rangle = \rho_0^2 - \frac{K}{2\pi^2} r^{-2} + A_3 \cos(2\pi \rho r) r^{-2K} + \cdots,
%\end{equation}
%where $A_3$ is a non-universal constant which depends on the microscopic properties of the individual model. It is clear then that for this case we must have $A_3\sim0$, such that the condensation peak does not depend on the density. Additionally, there is an underlying anti-ferromagnetic order, although with uniform density, which brings into the above expression a correction factor, $(-1)^r$, giving the Fourier transform a dominant phase at $ka=\pi$. 
%}

\section{Phase Separation}

There is still the question of whether the regimes where the two phases exist simultaneously are really lattice supersolids. In the main text we calculate the energy dispersion relations above the ground state in an attempt to detect supersolid signatures in the excitations, but there is also the question of phase separation. Will the CDW and PLL phases exist uniformly throughout the whole system or will there be distinct regions of one and separate regions of the other? In an attempt to answer this, we consider the $\rho<2$ PSS phase and compare the free energy, $\tilde{E} = E(\mu_1)/2 + E(\mu_2)/2$, with $E(\mu_0)$ such that the density, $\rho(\mu_1)/2 + \rho(\mu_2)/2 = \rho(\mu_0)$. Where we choose $\mu_0$ to be in the $\rho<2$ PSS phase, $\mu_1$, to be in the $\rho>1$ PLL phase and $\mu_2$ to be in the $\rho=2$ CDW phase. This quantifies if it is more energy favourable, for a system with a given total particle number, to have regions of lower and regions of higher density, or for it to have uniform density. The results of this analysis are that it is more favourable to have uniform density, indicating that phase separation in the $\rho<2$ PSS phase does not occur. However, finite size effects in a real experimental setting may alter this behaviour so future care must be taken.

\section{Adiabatic Preparation Scheme}

In cold-atom experiments, if the temperature is much smaller than the critical temperature for condensation then effectively the system is at zero temperature and we can model the dynamics as a pure state. The main consideration in the preparation of low energy eigenstates is in reducing the overall entropy of the many-body state. In adiabatic state preparation~\cite{PhysRevA.88.012334,Simon:2011aa}, this is achieved by first producing a low entropy initial state, in this case by projecting single atoms onto single sites, and ensuring that the state is the ground state of the initial Hamiltonian, in this case by having the onsite energies of populated sites at a much lower energy than the others. This also ensures that the trapped atoms have no dynamics. We then ramp the parameters of the lattice so as to create the final Hamiltonian that we are interested in. And if this ramp process is slow enough, so as not to induce unwanted heating effects, but fast enough so that decoherence effects can still be ignored then we can produce the desired low energy eigenstate with also a low entropy. 

We begin with atoms populating only particular unit-cells, where the number of populated unit cells is chosen to give the required density and are equally separated. On a populated unit cell, we have a single atom on each of the two sites, and we have the onsite energies of these populated sites at a significantly lower energy, $E_0/J=-\mu_0$ compared to the rest, $E=0$. We then ramp the energy of the populated sites to the value of the other sites, using the following exponential ramp,
\begin{equation} \label{Ramp_Equ}
E(t) = -\mu_0 \frac{e^{5(1-t/T)}-1}{e^5-1},
\end{equation}
where $T$ is the total time for the ramp. We ensure that the initial state is an eigenstate by beginning with a product state where we have the atoms localised on these initial sites and with all nearest-neighbour tunnelling amplitudes set to zero. We then linearly ramp all tunnelling terms from zero to one in a time $TJ=10$. The resulting state only has a small amplitude on sites around the ones with a lower onsite energy and has an overlap with the initial product state $>0.9$, but there are important phases in these new components which ensure that all tunnelling processes to sites at higher energy (although very highly suppressed in the product state) exactly cancel on the Creutz ladder geometry for the eigenstate. In the main text, we set the initial energy offset to $\mu_0=20 J$

Note that the correlations of the time-dependently produced state in the main text decay faster than the correlations of the ground state. This discrepancy arrises from the breakdown of the adiabatic principle due to the energy gap closing at the end of the ramp process. However the state that is produced is an eigenstate of the many-body Hamiltonian with a low energy variance, and from exact diagonalisation analysis of smaller systems, we find that the prepared state is the first excited state.

\end{widetext}

%%%%% REFERENCES
%\bibliographystyle{apsrev-title}
\bibliography{EXP_Creutz_Tex.bib}

\begin{thebibliography}{108}%
\makeatletter
\providecommand \@ifxundefined [1]{%
 \@ifx{#1\undefined}
}%
\providecommand \@ifnum [1]{%
 \ifnum #1\expandafter \@firstoftwo
 \else \expandafter \@secondoftwo
 \fi
}%
\providecommand \@ifx [1]{%
 \ifx #1\expandafter \@firstoftwo
 \else \expandafter \@secondoftwo
 \fi
}%
\providecommand \natexlab [1]{#1}%
\providecommand \enquote  [1]{``#1''}%
\providecommand \bibnamefont  [1]{#1}%
\providecommand \bibfnamefont [1]{#1}%
\providecommand \citenamefont [1]{#1}%
\providecommand \href@noop [0]{\@secondoftwo}%
\providecommand \href [0]{\begingroup \@sanitize@url \@href}%
\providecommand \@href[1]{\@@startlink{#1}\@@href}%
\providecommand \@@href[1]{\endgroup#1\@@endlink}%
\providecommand \@sanitize@url [0]{\catcode `\\12\catcode `\$12\catcode
  `\&12\catcode `\#12\catcode `\^12\catcode `\_12\catcode `\%12\relax}%
\providecommand \@@startlink[1]{}%
\providecommand \@@endlink[0]{}%
\providecommand \url  [0]{\begingroup\@sanitize@url \@url }%
\providecommand \@url [1]{\endgroup\@href {#1}{\urlprefix }}%
\providecommand \urlprefix  [0]{URL }%
\providecommand \Eprint [0]{\href }%
\providecommand \doibase [0]{http://dx.doi.org/}%
\providecommand \selectlanguage [0]{\@gobble}%
\providecommand \bibinfo  [0]{\@secondoftwo}%
\providecommand \bibfield  [0]{\@secondoftwo}%
\providecommand \translation [1]{[#1]}%
\providecommand \BibitemOpen [0]{}%
\providecommand \bibitemStop [0]{}%
\providecommand \bibitemNoStop [0]{.\EOS\space}%
\providecommand \EOS [0]{\spacefactor3000\relax}%
\providecommand \BibitemShut  [1]{\csname bibitem#1\endcsname}%
\let\auto@bib@innerbib\@empty
%</preamble>
\bibitem [{\citenamefont {Lin}\ \emph {et~al.}(2011)\citenamefont {Lin},
  \citenamefont {Jim{\'e}nez-Garc{\'\i}a},\ and\ \citenamefont
  {Spielman}}]{Lin:2011aa}%
  \BibitemOpen
  \bibfield  {author} {\bibinfo {author} {\bibfnamefont {Y.~J.}\ \bibnamefont
  {Lin}}, \bibinfo {author} {\bibfnamefont {K.}~\bibnamefont
  {Jim{\'e}nez-Garc{\'\i}a}}, \ and\ \bibinfo {author} {\bibfnamefont {I.~B.}\
  \bibnamefont {Spielman}},\ }\href {\doibase 10.1038/nature09887} {\bibfield
  {journal} {\bibinfo  {journal} {Nature}\ }\textbf {\bibinfo {volume} {471}},\
  \bibinfo {pages} {83} (\bibinfo {year} {2011})}\BibitemShut {NoStop}%
\bibitem [{\citenamefont {Struck}\ \emph {et~al.}(2011)\citenamefont {Struck},
  \citenamefont {{\"O}lschl{\"a}ger}, \citenamefont {Le~Targat}, \citenamefont
  {Soltan-Panahi}, \citenamefont {Eckardt}, \citenamefont {Lewenstein},
  \citenamefont {Windpassinger},\ and\ \citenamefont {Sengstock}}]{Struck996}%
  \BibitemOpen
  \bibfield  {author} {\bibinfo {author} {\bibfnamefont {J.}~\bibnamefont
  {Struck}}, \bibinfo {author} {\bibfnamefont {C.}~\bibnamefont
  {{\"O}lschl{\"a}ger}}, \bibinfo {author} {\bibfnamefont {R.}~\bibnamefont
  {Le~Targat}}, \bibinfo {author} {\bibfnamefont {P.}~\bibnamefont
  {Soltan-Panahi}}, \bibinfo {author} {\bibfnamefont {A.}~\bibnamefont
  {Eckardt}}, \bibinfo {author} {\bibfnamefont {M.}~\bibnamefont {Lewenstein}},
  \bibinfo {author} {\bibfnamefont {P.}~\bibnamefont {Windpassinger}}, \ and\
  \bibinfo {author} {\bibfnamefont {K.}~\bibnamefont {Sengstock}},\ }\href
  {\doibase 10.1126/science.1207239} {\bibfield  {journal} {\bibinfo  {journal}
  {Science}\ }\textbf {\bibinfo {volume} {333}},\ \bibinfo {pages} {996}
  (\bibinfo {year} {2011})}\BibitemShut {NoStop}%
\bibitem [{\citenamefont {Struck}\ \emph {et~al.}(2012)\citenamefont {Struck},
  \citenamefont {\"Olschl\"ager}, \citenamefont {Weinberg}, \citenamefont
  {Hauke}, \citenamefont {Simonet}, \citenamefont {Eckardt}, \citenamefont
  {Lewenstein}, \citenamefont {Sengstock},\ and\ \citenamefont
  {Windpassinger}}]{PhysRevLett.108.225304}%
  \BibitemOpen
  \bibfield  {author} {\bibinfo {author} {\bibfnamefont {J.}~\bibnamefont
  {Struck}}, \bibinfo {author} {\bibfnamefont {C.}~\bibnamefont
  {\"Olschl\"ager}}, \bibinfo {author} {\bibfnamefont {M.}~\bibnamefont
  {Weinberg}}, \bibinfo {author} {\bibfnamefont {P.}~\bibnamefont {Hauke}},
  \bibinfo {author} {\bibfnamefont {J.}~\bibnamefont {Simonet}}, \bibinfo
  {author} {\bibfnamefont {A.}~\bibnamefont {Eckardt}}, \bibinfo {author}
  {\bibfnamefont {M.}~\bibnamefont {Lewenstein}}, \bibinfo {author}
  {\bibfnamefont {K.}~\bibnamefont {Sengstock}}, \ and\ \bibinfo {author}
  {\bibfnamefont {P.}~\bibnamefont {Windpassinger}},\ }\href {\doibase
  10.1103/PhysRevLett.108.225304} {\bibfield  {journal} {\bibinfo  {journal}
  {Phys. Rev. Lett.}\ }\textbf {\bibinfo {volume} {108}},\ \bibinfo {pages}
  {225304} (\bibinfo {year} {2012})}\BibitemShut {NoStop}%
\bibitem [{\citenamefont {Jotzu}\ \emph {et~al.}(2014)\citenamefont {Jotzu},
  \citenamefont {Messer}, \citenamefont {Desbuquois}, \citenamefont {Lebrat},
  \citenamefont {Uehlinger}, \citenamefont {Greif},\ and\ \citenamefont
  {Esslinger}}]{Jotzu:2014aa}%
  \BibitemOpen
  \bibfield  {author} {\bibinfo {author} {\bibfnamefont {G.}~\bibnamefont
  {Jotzu}}, \bibinfo {author} {\bibfnamefont {M.}~\bibnamefont {Messer}},
  \bibinfo {author} {\bibfnamefont {R.}~\bibnamefont {Desbuquois}}, \bibinfo
  {author} {\bibfnamefont {M.}~\bibnamefont {Lebrat}}, \bibinfo {author}
  {\bibfnamefont {T.}~\bibnamefont {Uehlinger}}, \bibinfo {author}
  {\bibfnamefont {D.}~\bibnamefont {Greif}}, \ and\ \bibinfo {author}
  {\bibfnamefont {T.}~\bibnamefont {Esslinger}},\ }\href {\doibase
  10.1038/nature13915} {\bibfield  {journal} {\bibinfo  {journal} {Nature}\
  }\textbf {\bibinfo {volume} {515}},\ \bibinfo {pages} {237} (\bibinfo {year}
  {2014})}\BibitemShut {NoStop}%
\bibitem [{\citenamefont {Aidelsburger}\ \emph {et~al.}(2011)\citenamefont
  {Aidelsburger}, \citenamefont {Atala}, \citenamefont {Nascimb\`ene},
  \citenamefont {Trotzky}, \citenamefont {Chen},\ and\ \citenamefont
  {Bloch}}]{PhysRevLett.107.255301}%
  \BibitemOpen
  \bibfield  {author} {\bibinfo {author} {\bibfnamefont {M.}~\bibnamefont
  {Aidelsburger}}, \bibinfo {author} {\bibfnamefont {M.}~\bibnamefont {Atala}},
  \bibinfo {author} {\bibfnamefont {S.}~\bibnamefont {Nascimb\`ene}}, \bibinfo
  {author} {\bibfnamefont {S.}~\bibnamefont {Trotzky}}, \bibinfo {author}
  {\bibfnamefont {Y.-A.}\ \bibnamefont {Chen}}, \ and\ \bibinfo {author}
  {\bibfnamefont {I.}~\bibnamefont {Bloch}},\ }\href {\doibase
  10.1103/PhysRevLett.107.255301} {\bibfield  {journal} {\bibinfo  {journal}
  {Phys. Rev. Lett.}\ }\textbf {\bibinfo {volume} {107}},\ \bibinfo {pages}
  {255301} (\bibinfo {year} {2011})}\BibitemShut {NoStop}%
\bibitem [{\citenamefont {Miyake}\ \emph {et~al.}(2013)\citenamefont {Miyake},
  \citenamefont {Siviloglou}, \citenamefont {Kennedy}, \citenamefont {Burton},\
  and\ \citenamefont {Ketterle}}]{PhysRevLett.111.185302}%
  \BibitemOpen
  \bibfield  {author} {\bibinfo {author} {\bibfnamefont {H.}~\bibnamefont
  {Miyake}}, \bibinfo {author} {\bibfnamefont {G.~A.}\ \bibnamefont
  {Siviloglou}}, \bibinfo {author} {\bibfnamefont {C.~J.}\ \bibnamefont
  {Kennedy}}, \bibinfo {author} {\bibfnamefont {W.~C.}\ \bibnamefont {Burton}},
  \ and\ \bibinfo {author} {\bibfnamefont {W.}~\bibnamefont {Ketterle}},\
  }\href {\doibase 10.1103/PhysRevLett.111.185302} {\bibfield  {journal}
  {\bibinfo  {journal} {Phys. Rev. Lett.}\ }\textbf {\bibinfo {volume} {111}},\
  \bibinfo {pages} {185302} (\bibinfo {year} {2013})}\BibitemShut {NoStop}%
\bibitem [{\citenamefont {Fl{\"a}schner}\ \emph {et~al.}(2016)\citenamefont
  {Fl{\"a}schner}, \citenamefont {Rem}, \citenamefont {Tarnowski},
  \citenamefont {Vogel}, \citenamefont {L{\"u}hmann}, \citenamefont
  {Sengstock},\ and\ \citenamefont {Weitenberg}}]{Flaschner:2016aa}%
  \BibitemOpen
  \bibfield  {author} {\bibinfo {author} {\bibfnamefont {N.}~\bibnamefont
  {Fl{\"a}schner}}, \bibinfo {author} {\bibfnamefont {B.~S.}\ \bibnamefont
  {Rem}}, \bibinfo {author} {\bibfnamefont {M.}~\bibnamefont {Tarnowski}},
  \bibinfo {author} {\bibfnamefont {D.}~\bibnamefont {Vogel}}, \bibinfo
  {author} {\bibfnamefont {D.~S.}\ \bibnamefont {L{\"u}hmann}}, \bibinfo
  {author} {\bibfnamefont {K.}~\bibnamefont {Sengstock}}, \ and\ \bibinfo
  {author} {\bibfnamefont {C.}~\bibnamefont {Weitenberg}},\ }\href {\doibase
  10.1126/science.aad4568} {\bibfield  {journal} {\bibinfo  {journal}
  {Science}\ }\textbf {\bibinfo {volume} {352}},\ \bibinfo {pages} {1091}
  (\bibinfo {year} {2016})}\BibitemShut {NoStop}%
\bibitem [{\citenamefont {Tarnowski}\ \emph {et~al.}(2017)\citenamefont
  {Tarnowski}, \citenamefont {Nuske}, \citenamefont {Fl\"aschner},
  \citenamefont {Rem}, \citenamefont {Vogel}, \citenamefont {Freystatzky},
  \citenamefont {Sengstock}, \citenamefont {Mathey},\ and\ \citenamefont
  {Weitenberg}}]{PhysRevLett.118.240403}%
  \BibitemOpen
  \bibfield  {author} {\bibinfo {author} {\bibfnamefont {M.}~\bibnamefont
  {Tarnowski}}, \bibinfo {author} {\bibfnamefont {M.}~\bibnamefont {Nuske}},
  \bibinfo {author} {\bibfnamefont {N.}~\bibnamefont {Fl\"aschner}}, \bibinfo
  {author} {\bibfnamefont {B.}~\bibnamefont {Rem}}, \bibinfo {author}
  {\bibfnamefont {D.}~\bibnamefont {Vogel}}, \bibinfo {author} {\bibfnamefont
  {L.}~\bibnamefont {Freystatzky}}, \bibinfo {author} {\bibfnamefont
  {K.}~\bibnamefont {Sengstock}}, \bibinfo {author} {\bibfnamefont
  {L.}~\bibnamefont {Mathey}}, \ and\ \bibinfo {author} {\bibfnamefont
  {C.}~\bibnamefont {Weitenberg}},\ }\href {\doibase
  10.1103/PhysRevLett.118.240403} {\bibfield  {journal} {\bibinfo  {journal}
  {Phys. Rev. Lett.}\ }\textbf {\bibinfo {volume} {118}},\ \bibinfo {pages}
  {240403} (\bibinfo {year} {2017})}\BibitemShut {NoStop}%
\bibitem [{\citenamefont {Aidelsburger}(2018)}]{Aidelsburger:2018aa}%
  \BibitemOpen
  \bibfield  {author} {\bibinfo {author} {\bibfnamefont {M.}~\bibnamefont
  {Aidelsburger}},\ }\href {\doibase 10.1088/1361-6455/aac120} {\bibfield
  {journal} {\bibinfo  {journal} {Journal of Physics B: Atomic, Molecular and
  Optical Physics}\ }\textbf {\bibinfo {volume} {51}},\ \bibinfo {pages}
  {193001} (\bibinfo {year} {2018})}\BibitemShut {NoStop}%
\bibitem [{\citenamefont {Aidelsburger}\ \emph {et~al.}(2013)\citenamefont
  {Aidelsburger}, \citenamefont {Atala}, \citenamefont {Lohse}, \citenamefont
  {Barreiro}, \citenamefont {Paredes},\ and\ \citenamefont
  {Bloch}}]{PhysRevLett.111.185301}%
  \BibitemOpen
  \bibfield  {author} {\bibinfo {author} {\bibfnamefont {M.}~\bibnamefont
  {Aidelsburger}}, \bibinfo {author} {\bibfnamefont {M.}~\bibnamefont {Atala}},
  \bibinfo {author} {\bibfnamefont {M.}~\bibnamefont {Lohse}}, \bibinfo
  {author} {\bibfnamefont {J.~T.}\ \bibnamefont {Barreiro}}, \bibinfo {author}
  {\bibfnamefont {B.}~\bibnamefont {Paredes}}, \ and\ \bibinfo {author}
  {\bibfnamefont {I.}~\bibnamefont {Bloch}},\ }\href {\doibase
  10.1103/PhysRevLett.111.185301} {\bibfield  {journal} {\bibinfo  {journal}
  {Phys. Rev. Lett.}\ }\textbf {\bibinfo {volume} {111}},\ \bibinfo {pages}
  {185301} (\bibinfo {year} {2013})}\BibitemShut {NoStop}%
\bibitem [{\citenamefont {Kooi}\ \emph {et~al.}(2018)\citenamefont {Kooi},
  \citenamefont {Quelle}, \citenamefont {Beugeling},\ and\ \citenamefont
  {Morais~Smith}}]{PhysRevB.98.115124}%
  \BibitemOpen
  \bibfield  {author} {\bibinfo {author} {\bibfnamefont {S.~H.}\ \bibnamefont
  {Kooi}}, \bibinfo {author} {\bibfnamefont {A.}~\bibnamefont {Quelle}},
  \bibinfo {author} {\bibfnamefont {W.}~\bibnamefont {Beugeling}}, \ and\
  \bibinfo {author} {\bibfnamefont {C.}~\bibnamefont {Morais~Smith}},\ }\href
  {\doibase 10.1103/PhysRevB.98.115124} {\bibfield  {journal} {\bibinfo
  {journal} {Phys. Rev. B}\ }\textbf {\bibinfo {volume} {98}},\ \bibinfo
  {pages} {115124} (\bibinfo {year} {2018})}\BibitemShut {NoStop}%
\bibitem [{\citenamefont {Quelle}\ \emph {et~al.}(2017)\citenamefont {Quelle},
  \citenamefont {Weitenberg}, \citenamefont {Sengstock},\ and\ \citenamefont
  {Smith}}]{Quelle:2017aa}%
  \BibitemOpen
  \bibfield  {author} {\bibinfo {author} {\bibfnamefont {A.}~\bibnamefont
  {Quelle}}, \bibinfo {author} {\bibfnamefont {C.}~\bibnamefont {Weitenberg}},
  \bibinfo {author} {\bibfnamefont {K.}~\bibnamefont {Sengstock}}, \ and\
  \bibinfo {author} {\bibfnamefont {C.~M.}\ \bibnamefont {Smith}},\ }\href
  {\doibase 10.1088/1367-2630/aa8646} {\bibfield  {journal} {\bibinfo
  {journal} {New Journal of Physics}\ }\textbf {\bibinfo {volume} {19}},\
  \bibinfo {pages} {113010} (\bibinfo {year} {2017})}\BibitemShut {NoStop}%
\bibitem [{\citenamefont {Bansil}\ \emph {et~al.}(2016)\citenamefont {Bansil},
  \citenamefont {Lin},\ and\ \citenamefont {Das}}]{RevModPhys.88.021004}%
  \BibitemOpen
  \bibfield  {author} {\bibinfo {author} {\bibfnamefont {A.}~\bibnamefont
  {Bansil}}, \bibinfo {author} {\bibfnamefont {H.}~\bibnamefont {Lin}}, \ and\
  \bibinfo {author} {\bibfnamefont {T.}~\bibnamefont {Das}},\ }\href {\doibase
  10.1103/RevModPhys.88.021004} {\bibfield  {journal} {\bibinfo  {journal}
  {Rev. Mod. Phys.}\ }\textbf {\bibinfo {volume} {88}},\ \bibinfo {pages}
  {021004} (\bibinfo {year} {2016})}\BibitemShut {NoStop}%
\bibitem [{\citenamefont {Hasan}\ and\ \citenamefont
  {Kane}(2010)}]{RevModPhys.82.3045}%
  \BibitemOpen
  \bibfield  {author} {\bibinfo {author} {\bibfnamefont {M.~Z.}\ \bibnamefont
  {Hasan}}\ and\ \bibinfo {author} {\bibfnamefont {C.~L.}\ \bibnamefont
  {Kane}},\ }\href {\doibase 10.1103/RevModPhys.82.3045} {\bibfield  {journal}
  {\bibinfo  {journal} {Rev. Mod. Phys.}\ }\textbf {\bibinfo {volume} {82}},\
  \bibinfo {pages} {3045} (\bibinfo {year} {2010})}\BibitemShut {NoStop}%
\bibitem [{\citenamefont {Aidelsburger}\ \emph {et~al.}(2015)\citenamefont
  {Aidelsburger}, \citenamefont {Lohse}, \citenamefont {Schweizer},
  \citenamefont {Atala}, \citenamefont {Barreiro}, \citenamefont
  {Nascimb{\`e}ne}, \citenamefont {Cooper}, \citenamefont {Bloch},\ and\
  \citenamefont {Goldman}}]{Aidelsburger:2015aa}%
  \BibitemOpen
  \bibfield  {author} {\bibinfo {author} {\bibfnamefont {M.}~\bibnamefont
  {Aidelsburger}}, \bibinfo {author} {\bibfnamefont {M.}~\bibnamefont {Lohse}},
  \bibinfo {author} {\bibfnamefont {C.}~\bibnamefont {Schweizer}}, \bibinfo
  {author} {\bibfnamefont {M.}~\bibnamefont {Atala}}, \bibinfo {author}
  {\bibfnamefont {J.~T.}\ \bibnamefont {Barreiro}}, \bibinfo {author}
  {\bibfnamefont {S.}~\bibnamefont {Nascimb{\`e}ne}}, \bibinfo {author}
  {\bibfnamefont {N.~R.}\ \bibnamefont {Cooper}}, \bibinfo {author}
  {\bibfnamefont {I.}~\bibnamefont {Bloch}}, \ and\ \bibinfo {author}
  {\bibfnamefont {N.}~\bibnamefont {Goldman}},\ }\href {\doibase
  10.1038/nphys3171} {\bibfield  {journal} {\bibinfo  {journal} {Nature
  Physics}\ }\textbf {\bibinfo {volume} {11}},\ \bibinfo {pages} {162}
  (\bibinfo {year} {2015})}\BibitemShut {NoStop}%
\bibitem [{\citenamefont {Atala}\ \emph {et~al.}(2013)\citenamefont {Atala},
  \citenamefont {Aidelsburger}, \citenamefont {Barreiro}, \citenamefont
  {Abanin}, \citenamefont {Kitagawa}, \citenamefont {Demler},\ and\
  \citenamefont {Bloch}}]{Atala:2013aa}%
  \BibitemOpen
  \bibfield  {author} {\bibinfo {author} {\bibfnamefont {M.}~\bibnamefont
  {Atala}}, \bibinfo {author} {\bibfnamefont {M.}~\bibnamefont {Aidelsburger}},
  \bibinfo {author} {\bibfnamefont {J.~T.}\ \bibnamefont {Barreiro}}, \bibinfo
  {author} {\bibfnamefont {D.}~\bibnamefont {Abanin}}, \bibinfo {author}
  {\bibfnamefont {T.}~\bibnamefont {Kitagawa}}, \bibinfo {author}
  {\bibfnamefont {E.}~\bibnamefont {Demler}}, \ and\ \bibinfo {author}
  {\bibfnamefont {I.}~\bibnamefont {Bloch}},\ }\href
  {https://doi.org/10.1038/nphys2790} {\bibfield  {journal} {\bibinfo
  {journal} {Nature Physics}\ }\textbf {\bibinfo {volume} {9}},\ \bibinfo
  {pages} {795 EP } (\bibinfo {year} {2013})}\BibitemShut {NoStop}%
\bibitem [{\citenamefont {Asteria}\ \emph {et~al.}(2019)\citenamefont
  {Asteria}, \citenamefont {Tran}, \citenamefont {Ozawa}, \citenamefont
  {Tarnowski}, \citenamefont {Rem}, \citenamefont {Fl{\"a}schner},
  \citenamefont {Sengstock}, \citenamefont {Goldman},\ and\ \citenamefont
  {Weitenberg}}]{Asteria:2019aa}%
  \BibitemOpen
  \bibfield  {author} {\bibinfo {author} {\bibfnamefont {L.}~\bibnamefont
  {Asteria}}, \bibinfo {author} {\bibfnamefont {D.~T.}\ \bibnamefont {Tran}},
  \bibinfo {author} {\bibfnamefont {T.}~\bibnamefont {Ozawa}}, \bibinfo
  {author} {\bibfnamefont {M.}~\bibnamefont {Tarnowski}}, \bibinfo {author}
  {\bibfnamefont {B.~S.}\ \bibnamefont {Rem}}, \bibinfo {author} {\bibfnamefont
  {N.}~\bibnamefont {Fl{\"a}schner}}, \bibinfo {author} {\bibfnamefont
  {K.}~\bibnamefont {Sengstock}}, \bibinfo {author} {\bibfnamefont
  {N.}~\bibnamefont {Goldman}}, \ and\ \bibinfo {author} {\bibfnamefont
  {C.}~\bibnamefont {Weitenberg}},\ }\href {\doibase 10.1038/s41567-019-0417-8}
  {\bibfield  {journal} {\bibinfo  {journal} {Nature Physics}\ }\textbf
  {\bibinfo {volume} {15}},\ \bibinfo {pages} {449} (\bibinfo {year}
  {2019})}\BibitemShut {NoStop}%
\bibitem [{\citenamefont {Tarnowski}\ \emph {et~al.}(2019)\citenamefont
  {Tarnowski}, \citenamefont {{\"U}nal}, \citenamefont {Fl{\"a}schner},
  \citenamefont {Rem}, \citenamefont {Eckardt}, \citenamefont {Sengstock},\
  and\ \citenamefont {Weitenberg}}]{Tarnowski:2019aa}%
  \BibitemOpen
  \bibfield  {author} {\bibinfo {author} {\bibfnamefont {M.}~\bibnamefont
  {Tarnowski}}, \bibinfo {author} {\bibfnamefont {F.~N.}\ \bibnamefont
  {{\"U}nal}}, \bibinfo {author} {\bibfnamefont {N.}~\bibnamefont
  {Fl{\"a}schner}}, \bibinfo {author} {\bibfnamefont {B.~S.}\ \bibnamefont
  {Rem}}, \bibinfo {author} {\bibfnamefont {A.}~\bibnamefont {Eckardt}},
  \bibinfo {author} {\bibfnamefont {K.}~\bibnamefont {Sengstock}}, \ and\
  \bibinfo {author} {\bibfnamefont {C.}~\bibnamefont {Weitenberg}},\ }\href
  {\doibase 10.1038/s41467-019-09668-y} {\bibfield  {journal} {\bibinfo
  {journal} {Nature Communications}\ }\textbf {\bibinfo {volume} {10}},\
  \bibinfo {pages} {1728} (\bibinfo {year} {2019})}\BibitemShut {NoStop}%
\bibitem [{\citenamefont {Wimmer}\ \emph {et~al.}(2017)\citenamefont {Wimmer},
  \citenamefont {Price}, \citenamefont {Carusotto},\ and\ \citenamefont
  {Peschel}}]{Wimmer:2017aa}%
  \BibitemOpen
  \bibfield  {author} {\bibinfo {author} {\bibfnamefont {M.}~\bibnamefont
  {Wimmer}}, \bibinfo {author} {\bibfnamefont {H.~M.}\ \bibnamefont {Price}},
  \bibinfo {author} {\bibfnamefont {I.}~\bibnamefont {Carusotto}}, \ and\
  \bibinfo {author} {\bibfnamefont {U.}~\bibnamefont {Peschel}},\ }\href
  {\doibase 10.1038/nphys4050} {\bibfield  {journal} {\bibinfo  {journal}
  {Nature Physics}\ }\textbf {\bibinfo {volume} {13}},\ \bibinfo {pages} {545}
  (\bibinfo {year} {2017})}\BibitemShut {NoStop}%
\bibitem [{\citenamefont {Cooper}\ \emph {et~al.}(2019)\citenamefont {Cooper},
  \citenamefont {Dalibard},\ and\ \citenamefont
  {Spielman}}]{RevModPhys.91.015005}%
  \BibitemOpen
  \bibfield  {author} {\bibinfo {author} {\bibfnamefont {N.~R.}\ \bibnamefont
  {Cooper}}, \bibinfo {author} {\bibfnamefont {J.}~\bibnamefont {Dalibard}}, \
  and\ \bibinfo {author} {\bibfnamefont {I.~B.}\ \bibnamefont {Spielman}},\
  }\href {\doibase 10.1103/RevModPhys.91.015005} {\bibfield  {journal}
  {\bibinfo  {journal} {Rev. Mod. Phys.}\ }\textbf {\bibinfo {volume} {91}},\
  \bibinfo {pages} {015005} (\bibinfo {year} {2019})}\BibitemShut {NoStop}%
\bibitem [{\citenamefont {Cooper}\ and\ \citenamefont
  {Dalibard}(2013)}]{PhysRevLett.110.185301}%
  \BibitemOpen
  \bibfield  {author} {\bibinfo {author} {\bibfnamefont {N.~R.}\ \bibnamefont
  {Cooper}}\ and\ \bibinfo {author} {\bibfnamefont {J.}~\bibnamefont
  {Dalibard}},\ }\href {\doibase 10.1103/PhysRevLett.110.185301} {\bibfield
  {journal} {\bibinfo  {journal} {Phys. Rev. Lett.}\ }\textbf {\bibinfo
  {volume} {110}},\ \bibinfo {pages} {185301} (\bibinfo {year}
  {2013})}\BibitemShut {NoStop}%
\bibitem [{\citenamefont {Grushin}\ \emph {et~al.}(2014)\citenamefont
  {Grushin}, \citenamefont {G\'omez-Le\'on},\ and\ \citenamefont
  {Neupert}}]{PhysRevLett.112.156801}%
  \BibitemOpen
  \bibfield  {author} {\bibinfo {author} {\bibfnamefont {A.~G.}\ \bibnamefont
  {Grushin}}, \bibinfo {author} {\bibfnamefont {A.}~\bibnamefont
  {G\'omez-Le\'on}}, \ and\ \bibinfo {author} {\bibfnamefont {T.}~\bibnamefont
  {Neupert}},\ }\href {\doibase 10.1103/PhysRevLett.112.156801} {\bibfield
  {journal} {\bibinfo  {journal} {Phys. Rev. Lett.}\ }\textbf {\bibinfo
  {volume} {112}},\ \bibinfo {pages} {156801} (\bibinfo {year}
  {2014})}\BibitemShut {NoStop}%
\bibitem [{\citenamefont {Anisimovas}\ \emph {et~al.}(2015)\citenamefont
  {Anisimovas}, \citenamefont {\ifmmode~\check{Z}\else \v{Z}\fi{}labys},
  \citenamefont {Anderson}, \citenamefont {Juzeli\ifmmode~\bar{u}\else
  \={u}\fi{}nas},\ and\ \citenamefont {Eckardt}}]{PhysRevB.91.245135}%
  \BibitemOpen
  \bibfield  {author} {\bibinfo {author} {\bibfnamefont {E.}~\bibnamefont
  {Anisimovas}}, \bibinfo {author} {\bibfnamefont {G.}~\bibnamefont
  {\ifmmode~\check{Z}\else \v{Z}\fi{}labys}}, \bibinfo {author} {\bibfnamefont
  {B.~M.}\ \bibnamefont {Anderson}}, \bibinfo {author} {\bibfnamefont
  {G.}~\bibnamefont {Juzeli\ifmmode~\bar{u}\else \={u}\fi{}nas}}, \ and\
  \bibinfo {author} {\bibfnamefont {A.}~\bibnamefont {Eckardt}},\ }\href
  {\doibase 10.1103/PhysRevB.91.245135} {\bibfield  {journal} {\bibinfo
  {journal} {Phys. Rev. B}\ }\textbf {\bibinfo {volume} {91}},\ \bibinfo
  {pages} {245135} (\bibinfo {year} {2015})}\BibitemShut {NoStop}%
\bibitem [{\citenamefont {Aidelsburger}\ \emph {et~al.}(2018)\citenamefont
  {Aidelsburger}, \citenamefont {Nascimbene},\ and\ \citenamefont
  {Goldman}}]{Aidelsburger:2018ab}%
  \BibitemOpen
  \bibfield  {author} {\bibinfo {author} {\bibfnamefont {M.}~\bibnamefont
  {Aidelsburger}}, \bibinfo {author} {\bibfnamefont {S.}~\bibnamefont
  {Nascimbene}}, \ and\ \bibinfo {author} {\bibfnamefont {N.}~\bibnamefont
  {Goldman}},\ }\bibfield  {booktitle} {\emph {\bibinfo {booktitle} {Quantum
  simulation / Simulation quantique}},\ }\href {\doibase
  https://doi.org/10.1016/j.crhy.2018.03.002} {\bibfield  {journal} {\bibinfo
  {journal} {Comptes Rendus Physique}\ }\textbf {\bibinfo {volume} {19}},\
  \bibinfo {pages} {394} (\bibinfo {year} {2018})}\BibitemShut {NoStop}%
\bibitem [{\citenamefont {Barbiero}\ \emph {et~al.}(2018)\citenamefont
  {Barbiero}, \citenamefont {Santos},\ and\ \citenamefont
  {Goldman}}]{PhysRevB.97.201115}%
  \BibitemOpen
  \bibfield  {author} {\bibinfo {author} {\bibfnamefont {L.}~\bibnamefont
  {Barbiero}}, \bibinfo {author} {\bibfnamefont {L.}~\bibnamefont {Santos}}, \
  and\ \bibinfo {author} {\bibfnamefont {N.}~\bibnamefont {Goldman}},\ }\href
  {\doibase 10.1103/PhysRevB.97.201115} {\bibfield  {journal} {\bibinfo
  {journal} {Phys. Rev. B}\ }\textbf {\bibinfo {volume} {97}},\ \bibinfo
  {pages} {201115} (\bibinfo {year} {2018})}\BibitemShut {NoStop}%
\bibitem [{\citenamefont {Cardarelli}\ \emph {et~al.}(2017)\citenamefont
  {Cardarelli}, \citenamefont {Greschner},\ and\ \citenamefont
  {Santos}}]{PhysRevLett.119.180402}%
  \BibitemOpen
  \bibfield  {author} {\bibinfo {author} {\bibfnamefont {L.}~\bibnamefont
  {Cardarelli}}, \bibinfo {author} {\bibfnamefont {S.}~\bibnamefont
  {Greschner}}, \ and\ \bibinfo {author} {\bibfnamefont {L.}~\bibnamefont
  {Santos}},\ }\href {\doibase 10.1103/PhysRevLett.119.180402} {\bibfield
  {journal} {\bibinfo  {journal} {Phys. Rev. Lett.}\ }\textbf {\bibinfo
  {volume} {119}},\ \bibinfo {pages} {180402} (\bibinfo {year}
  {2017})}\BibitemShut {NoStop}%
\bibitem [{\citenamefont {Ravent\'os}\ \emph {et~al.}(2016)\citenamefont
  {Ravent\'os}, \citenamefont {Gra\ss{}}, \citenamefont {Juli\'a-D\'{\i}az},
  \citenamefont {Santos},\ and\ \citenamefont
  {Lewenstein}}]{PhysRevA.93.033605}%
  \BibitemOpen
  \bibfield  {author} {\bibinfo {author} {\bibfnamefont {D.}~\bibnamefont
  {Ravent\'os}}, \bibinfo {author} {\bibfnamefont {T.}~\bibnamefont
  {Gra\ss{}}}, \bibinfo {author} {\bibfnamefont {B.}~\bibnamefont
  {Juli\'a-D\'{\i}az}}, \bibinfo {author} {\bibfnamefont {L.}~\bibnamefont
  {Santos}}, \ and\ \bibinfo {author} {\bibfnamefont {M.}~\bibnamefont
  {Lewenstein}},\ }\href {\doibase 10.1103/PhysRevA.93.033605} {\bibfield
  {journal} {\bibinfo  {journal} {Phys. Rev. A}\ }\textbf {\bibinfo {volume}
  {93}},\ \bibinfo {pages} {033605} (\bibinfo {year} {2016})}\BibitemShut
  {NoStop}%
\bibitem [{\citenamefont {Greschner}\ \emph {et~al.}(2015)\citenamefont
  {Greschner}, \citenamefont {Huerga}, \citenamefont {Sun}, \citenamefont
  {Poletti},\ and\ \citenamefont {Santos}}]{PhysRevB.92.115120}%
  \BibitemOpen
  \bibfield  {author} {\bibinfo {author} {\bibfnamefont {S.}~\bibnamefont
  {Greschner}}, \bibinfo {author} {\bibfnamefont {D.}~\bibnamefont {Huerga}},
  \bibinfo {author} {\bibfnamefont {G.}~\bibnamefont {Sun}}, \bibinfo {author}
  {\bibfnamefont {D.}~\bibnamefont {Poletti}}, \ and\ \bibinfo {author}
  {\bibfnamefont {L.}~\bibnamefont {Santos}},\ }\href {\doibase
  10.1103/PhysRevB.92.115120} {\bibfield  {journal} {\bibinfo  {journal} {Phys.
  Rev. B}\ }\textbf {\bibinfo {volume} {92}},\ \bibinfo {pages} {115120}
  (\bibinfo {year} {2015})}\BibitemShut {NoStop}%
\bibitem [{\citenamefont {Deng}\ and\ \citenamefont
  {Santos}(2014)}]{PhysRevA.89.033632}%
  \BibitemOpen
  \bibfield  {author} {\bibinfo {author} {\bibfnamefont {X.}~\bibnamefont
  {Deng}}\ and\ \bibinfo {author} {\bibfnamefont {L.}~\bibnamefont {Santos}},\
  }\href {\doibase 10.1103/PhysRevA.89.033632} {\bibfield  {journal} {\bibinfo
  {journal} {Phys. Rev. A}\ }\textbf {\bibinfo {volume} {89}},\ \bibinfo
  {pages} {033632} (\bibinfo {year} {2014})}\BibitemShut {NoStop}%
\bibitem [{\citenamefont {Di~Liberto}\ \emph
  {et~al.}(2016{\natexlab{a}})\citenamefont {Di~Liberto}, \citenamefont
  {Hemmerich},\ and\ \citenamefont {Morais~Smith}}]{PhysRevLett.117.163001}%
  \BibitemOpen
  \bibfield  {author} {\bibinfo {author} {\bibfnamefont {M.}~\bibnamefont
  {Di~Liberto}}, \bibinfo {author} {\bibfnamefont {A.}~\bibnamefont
  {Hemmerich}}, \ and\ \bibinfo {author} {\bibfnamefont {C.}~\bibnamefont
  {Morais~Smith}},\ }\href {\doibase 10.1103/PhysRevLett.117.163001} {\bibfield
   {journal} {\bibinfo  {journal} {Phys. Rev. Lett.}\ }\textbf {\bibinfo
  {volume} {117}},\ \bibinfo {pages} {163001} (\bibinfo {year}
  {2016}{\natexlab{a}})}\BibitemShut {NoStop}%
\bibitem [{\citenamefont {Leykam}\ \emph {et~al.}(2018)\citenamefont {Leykam},
  \citenamefont {Andreanov},\ and\ \citenamefont
  {Flach}}]{doi:10.1080/23746149.2018.1473052}%
  \BibitemOpen
  \bibfield  {author} {\bibinfo {author} {\bibfnamefont {D.}~\bibnamefont
  {Leykam}}, \bibinfo {author} {\bibfnamefont {A.}~\bibnamefont {Andreanov}}, \
  and\ \bibinfo {author} {\bibfnamefont {S.}~\bibnamefont {Flach}},\ }\href
  {\doibase 10.1080/23746149.2018.1473052} {\bibfield  {journal} {\bibinfo
  {journal} {Advances in Physics: X}\ }\textbf {\bibinfo {volume} {3}},\
  \bibinfo {pages} {1473052} (\bibinfo {year} {2018})}\BibitemShut {NoStop}%
\bibitem [{\citenamefont {Liu}\ \emph {et~al.}(2014)\citenamefont {Liu},
  \citenamefont {Liu},\ and\ \citenamefont {Wu}}]{Liu_2014}%
  \BibitemOpen
  \bibfield  {author} {\bibinfo {author} {\bibfnamefont {Z.}~\bibnamefont
  {Liu}}, \bibinfo {author} {\bibfnamefont {F.}~\bibnamefont {Liu}}, \ and\
  \bibinfo {author} {\bibfnamefont {Y.-S.}\ \bibnamefont {Wu}},\ }\href
  {\doibase 10.1088/1674-1056/23/7/077308} {\bibfield  {journal} {\bibinfo
  {journal} {Chinese Physics B}\ }\textbf {\bibinfo {volume} {23}},\ \bibinfo
  {pages} {077308} (\bibinfo {year} {2014})}\BibitemShut {NoStop}%
\bibitem [{\citenamefont {Vidal}\ \emph {et~al.}(1998)\citenamefont {Vidal},
  \citenamefont {Mosseri},\ and\ \citenamefont {Dou\ifmmode~\mbox{\c{c}}\else
  \c{c}\fi{}ot}}]{PhysRevLett.81.5888}%
  \BibitemOpen
  \bibfield  {author} {\bibinfo {author} {\bibfnamefont {J.}~\bibnamefont
  {Vidal}}, \bibinfo {author} {\bibfnamefont {R.}~\bibnamefont {Mosseri}}, \
  and\ \bibinfo {author} {\bibfnamefont {B.}~\bibnamefont
  {Dou\ifmmode~\mbox{\c{c}}\else \c{c}\fi{}ot}},\ }\href {\doibase
  10.1103/PhysRevLett.81.5888} {\bibfield  {journal} {\bibinfo  {journal}
  {Phys. Rev. Lett.}\ }\textbf {\bibinfo {volume} {81}},\ \bibinfo {pages}
  {5888} (\bibinfo {year} {1998})}\BibitemShut {NoStop}%
\bibitem [{\citenamefont {Mukherjee}\ and\ \citenamefont
  {Thomson}(2015)}]{Mukherjee:15}%
  \BibitemOpen
  \bibfield  {author} {\bibinfo {author} {\bibfnamefont {S.}~\bibnamefont
  {Mukherjee}}\ and\ \bibinfo {author} {\bibfnamefont {R.~R.}\ \bibnamefont
  {Thomson}},\ }\href {\doibase 10.1364/OL.40.005443} {\bibfield  {journal}
  {\bibinfo  {journal} {Opt. Lett.}\ }\textbf {\bibinfo {volume} {40}},\
  \bibinfo {pages} {5443} (\bibinfo {year} {2015})}\BibitemShut {NoStop}%
\bibitem [{\citenamefont {Takayoshi}\ \emph {et~al.}(2013)\citenamefont
  {Takayoshi}, \citenamefont {Katsura}, \citenamefont {Watanabe},\ and\
  \citenamefont {Aoki}}]{PhysRevA.88.063613}%
  \BibitemOpen
  \bibfield  {author} {\bibinfo {author} {\bibfnamefont {S.}~\bibnamefont
  {Takayoshi}}, \bibinfo {author} {\bibfnamefont {H.}~\bibnamefont {Katsura}},
  \bibinfo {author} {\bibfnamefont {N.}~\bibnamefont {Watanabe}}, \ and\
  \bibinfo {author} {\bibfnamefont {H.}~\bibnamefont {Aoki}},\ }\href {\doibase
  10.1103/PhysRevA.88.063613} {\bibfield  {journal} {\bibinfo  {journal} {Phys.
  Rev. A}\ }\textbf {\bibinfo {volume} {88}},\ \bibinfo {pages} {063613}
  (\bibinfo {year} {2013})}\BibitemShut {NoStop}%
\bibitem [{\citenamefont {Tovmasyan}\ \emph {et~al.}(2013)\citenamefont
  {Tovmasyan}, \citenamefont {van Nieuwenburg},\ and\ \citenamefont
  {Huber}}]{PhysRevB.88.220510}%
  \BibitemOpen
  \bibfield  {author} {\bibinfo {author} {\bibfnamefont {M.}~\bibnamefont
  {Tovmasyan}}, \bibinfo {author} {\bibfnamefont {E.~P.~L.}\ \bibnamefont {van
  Nieuwenburg}}, \ and\ \bibinfo {author} {\bibfnamefont {S.~D.}\ \bibnamefont
  {Huber}},\ }\href {\doibase 10.1103/PhysRevB.88.220510} {\bibfield  {journal}
  {\bibinfo  {journal} {Phys. Rev. B}\ }\textbf {\bibinfo {volume} {88}},\
  \bibinfo {pages} {220510} (\bibinfo {year} {2013})}\BibitemShut {NoStop}%
\bibitem [{\citenamefont {M\"oller}\ and\ \citenamefont
  {Cooper}(2012)}]{PhysRevLett.108.045306}%
  \BibitemOpen
  \bibfield  {author} {\bibinfo {author} {\bibfnamefont {G.}~\bibnamefont
  {M\"oller}}\ and\ \bibinfo {author} {\bibfnamefont {N.~R.}\ \bibnamefont
  {Cooper}},\ }\href {\doibase 10.1103/PhysRevLett.108.045306} {\bibfield
  {journal} {\bibinfo  {journal} {Phys. Rev. Lett.}\ }\textbf {\bibinfo
  {volume} {108}},\ \bibinfo {pages} {045306} (\bibinfo {year}
  {2012})}\BibitemShut {NoStop}%
\bibitem [{\citenamefont {Huber}\ and\ \citenamefont
  {Altman}(2010)}]{PhysRevB.82.184502}%
  \BibitemOpen
  \bibfield  {author} {\bibinfo {author} {\bibfnamefont {S.~D.}\ \bibnamefont
  {Huber}}\ and\ \bibinfo {author} {\bibfnamefont {E.}~\bibnamefont {Altman}},\
  }\href {\doibase 10.1103/PhysRevB.82.184502} {\bibfield  {journal} {\bibinfo
  {journal} {Phys. Rev. B}\ }\textbf {\bibinfo {volume} {82}},\ \bibinfo
  {pages} {184502} (\bibinfo {year} {2010})}\BibitemShut {NoStop}%
\bibitem [{\citenamefont {Huhtinen}\ \emph {et~al.}(2018)\citenamefont
  {Huhtinen}, \citenamefont {Tylutki}, \citenamefont {Kumar}, \citenamefont
  {Vanhala}, \citenamefont {Peotta},\ and\ \citenamefont
  {T\"orm\"a}}]{PhysRevB.97.214503}%
  \BibitemOpen
  \bibfield  {author} {\bibinfo {author} {\bibfnamefont {K.-E.}\ \bibnamefont
  {Huhtinen}}, \bibinfo {author} {\bibfnamefont {M.}~\bibnamefont {Tylutki}},
  \bibinfo {author} {\bibfnamefont {P.}~\bibnamefont {Kumar}}, \bibinfo
  {author} {\bibfnamefont {T.~I.}\ \bibnamefont {Vanhala}}, \bibinfo {author}
  {\bibfnamefont {S.}~\bibnamefont {Peotta}}, \ and\ \bibinfo {author}
  {\bibfnamefont {P.}~\bibnamefont {T\"orm\"a}},\ }\href {\doibase
  10.1103/PhysRevB.97.214503} {\bibfield  {journal} {\bibinfo  {journal} {Phys.
  Rev. B}\ }\textbf {\bibinfo {volume} {97}},\ \bibinfo {pages} {214503}
  (\bibinfo {year} {2018})}\BibitemShut {NoStop}%
\bibitem [{\citenamefont {Tovmasyan}\ \emph {et~al.}(2018)\citenamefont
  {Tovmasyan}, \citenamefont {Peotta}, \citenamefont {Liang}, \citenamefont
  {T\"orm\"a},\ and\ \citenamefont {Huber}}]{PhysRevB.98.134513}%
  \BibitemOpen
  \bibfield  {author} {\bibinfo {author} {\bibfnamefont {M.}~\bibnamefont
  {Tovmasyan}}, \bibinfo {author} {\bibfnamefont {S.}~\bibnamefont {Peotta}},
  \bibinfo {author} {\bibfnamefont {L.}~\bibnamefont {Liang}}, \bibinfo
  {author} {\bibfnamefont {P.}~\bibnamefont {T\"orm\"a}}, \ and\ \bibinfo
  {author} {\bibfnamefont {S.~D.}\ \bibnamefont {Huber}},\ }\href {\doibase
  10.1103/PhysRevB.98.134513} {\bibfield  {journal} {\bibinfo  {journal} {Phys.
  Rev. B}\ }\textbf {\bibinfo {volume} {98}},\ \bibinfo {pages} {134513}
  (\bibinfo {year} {2018})}\BibitemShut {NoStop}%
\bibitem [{\citenamefont {Nunes}\ and\ \citenamefont
  {Morais~Smith}(2020)}]{MSmith}%
  \BibitemOpen
  \bibfield  {author} {\bibinfo {author} {\bibfnamefont {L.}~\bibnamefont
  {Nunes}}\ and\ \bibinfo {author} {\bibfnamefont {C.}~\bibnamefont
  {Morais~Smith}},\ }\href@noop {} {\bibfield  {journal} {\bibinfo  {journal}
  {arXiv:2003.09243}\ } (\bibinfo {year} {2020})}\BibitemShut {NoStop}%
\bibitem [{\citenamefont {Kumar}\ \emph {et~al.}(2020)\citenamefont {Kumar},
  \citenamefont {Peotta}, \citenamefont {Takasu}, \citenamefont {Takahashi},\
  and\ \citenamefont {T\"orm\"a}}]{PTorma}%
  \BibitemOpen
  \bibfield  {author} {\bibinfo {author} {\bibfnamefont {P.}~\bibnamefont
  {Kumar}}, \bibinfo {author} {\bibfnamefont {S.}~\bibnamefont {Peotta}},
  \bibinfo {author} {\bibfnamefont {Y.}~\bibnamefont {Takasu}}, \bibinfo
  {author} {\bibfnamefont {Y.}~\bibnamefont {Takahashi}}, \ and\ \bibinfo
  {author} {\bibfnamefont {P.}~\bibnamefont {T\"orm\"a}},\ }\href@noop {}
  {\bibfield  {journal} {\bibinfo  {journal} {arXiv:2005.05457}\ } (\bibinfo
  {year} {2020})}\BibitemShut {NoStop}%
\bibitem [{\citenamefont {Pelegr\'{\i}}\ \emph
  {et~al.}(2019{\natexlab{a}})\citenamefont {Pelegr\'{\i}}, \citenamefont
  {Marques}, \citenamefont {Dias}, \citenamefont {Daley}, \citenamefont
  {Mompart},\ and\ \citenamefont {Ahufinger}}]{PhysRevA.99.023613}%
  \BibitemOpen
  \bibfield  {author} {\bibinfo {author} {\bibfnamefont {G.}~\bibnamefont
  {Pelegr\'{\i}}}, \bibinfo {author} {\bibfnamefont {A.~M.}\ \bibnamefont
  {Marques}}, \bibinfo {author} {\bibfnamefont {R.~G.}\ \bibnamefont {Dias}},
  \bibinfo {author} {\bibfnamefont {A.~J.}\ \bibnamefont {Daley}}, \bibinfo
  {author} {\bibfnamefont {J.}~\bibnamefont {Mompart}}, \ and\ \bibinfo
  {author} {\bibfnamefont {V.}~\bibnamefont {Ahufinger}},\ }\href {\doibase
  10.1103/PhysRevA.99.023613} {\bibfield  {journal} {\bibinfo  {journal} {Phys.
  Rev. A}\ }\textbf {\bibinfo {volume} {99}},\ \bibinfo {pages} {023613}
  (\bibinfo {year} {2019}{\natexlab{a}})}\BibitemShut {NoStop}%
\bibitem [{\citenamefont {Pelegr\'{\i}}\ \emph
  {et~al.}(2019{\natexlab{b}})\citenamefont {Pelegr\'{\i}}, \citenamefont
  {Marques}, \citenamefont {Dias}, \citenamefont {Daley}, \citenamefont
  {Ahufinger},\ and\ \citenamefont {Mompart}}]{PhysRevA.99.023612}%
  \BibitemOpen
  \bibfield  {author} {\bibinfo {author} {\bibfnamefont {G.}~\bibnamefont
  {Pelegr\'{\i}}}, \bibinfo {author} {\bibfnamefont {A.~M.}\ \bibnamefont
  {Marques}}, \bibinfo {author} {\bibfnamefont {R.~G.}\ \bibnamefont {Dias}},
  \bibinfo {author} {\bibfnamefont {A.~J.}\ \bibnamefont {Daley}}, \bibinfo
  {author} {\bibfnamefont {V.}~\bibnamefont {Ahufinger}}, \ and\ \bibinfo
  {author} {\bibfnamefont {J.}~\bibnamefont {Mompart}},\ }\href {\doibase
  10.1103/PhysRevA.99.023612} {\bibfield  {journal} {\bibinfo  {journal} {Phys.
  Rev. A}\ }\textbf {\bibinfo {volume} {99}},\ \bibinfo {pages} {023612}
  (\bibinfo {year} {2019}{\natexlab{b}})}\BibitemShut {NoStop}%
\bibitem [{\citenamefont {J\"unemann}\ \emph {et~al.}(2017)\citenamefont
  {J\"unemann}, \citenamefont {Piga}, \citenamefont {Ran}, \citenamefont
  {Lewenstein}, \citenamefont {Rizzi},\ and\ \citenamefont
  {Bermudez}}]{PhysRevX.7.031057}%
  \BibitemOpen
  \bibfield  {author} {\bibinfo {author} {\bibfnamefont {J.}~\bibnamefont
  {J\"unemann}}, \bibinfo {author} {\bibfnamefont {A.}~\bibnamefont {Piga}},
  \bibinfo {author} {\bibfnamefont {S.-J.}\ \bibnamefont {Ran}}, \bibinfo
  {author} {\bibfnamefont {M.}~\bibnamefont {Lewenstein}}, \bibinfo {author}
  {\bibfnamefont {M.}~\bibnamefont {Rizzi}}, \ and\ \bibinfo {author}
  {\bibfnamefont {A.}~\bibnamefont {Bermudez}},\ }\href {\doibase
  10.1103/PhysRevX.7.031057} {\bibfield  {journal} {\bibinfo  {journal} {Phys.
  Rev. X}\ }\textbf {\bibinfo {volume} {7}},\ \bibinfo {pages} {031057}
  (\bibinfo {year} {2017})}\BibitemShut {NoStop}%
\bibitem [{\citenamefont {T\"orm\"a}\ \emph {et~al.}(2018)\citenamefont
  {T\"orm\"a}, \citenamefont {Liang},\ and\ \citenamefont
  {Peotta}}]{PhysRevB.98.220511}%
  \BibitemOpen
  \bibfield  {author} {\bibinfo {author} {\bibfnamefont {P.}~\bibnamefont
  {T\"orm\"a}}, \bibinfo {author} {\bibfnamefont {L.}~\bibnamefont {Liang}}, \
  and\ \bibinfo {author} {\bibfnamefont {S.}~\bibnamefont {Peotta}},\ }\href
  {\doibase 10.1103/PhysRevB.98.220511} {\bibfield  {journal} {\bibinfo
  {journal} {Phys. Rev. B}\ }\textbf {\bibinfo {volume} {98}},\ \bibinfo
  {pages} {220511} (\bibinfo {year} {2018})}\BibitemShut {NoStop}%
\bibitem [{\citenamefont {Mishra}\ \emph {et~al.}(2016)\citenamefont {Mishra},
  \citenamefont {Greschner},\ and\ \citenamefont {Santos}}]{Mishra_2016}%
  \BibitemOpen
  \bibfield  {author} {\bibinfo {author} {\bibfnamefont {T.}~\bibnamefont
  {Mishra}}, \bibinfo {author} {\bibfnamefont {S.}~\bibnamefont {Greschner}}, \
  and\ \bibinfo {author} {\bibfnamefont {L.}~\bibnamefont {Santos}},\ }\href
  {\doibase 10.1088/1367-2630/18/4/045016} {\bibfield  {journal} {\bibinfo
  {journal} {New Journal of Physics}\ }\textbf {\bibinfo {volume} {18}},\
  \bibinfo {pages} {045016} (\bibinfo {year} {2016})}\BibitemShut {NoStop}%
\bibitem [{\citenamefont {Tylutki}\ and\ \citenamefont
  {T\"orm\"a}(2018)}]{PhysRevB.98.094513}%
  \BibitemOpen
  \bibfield  {author} {\bibinfo {author} {\bibfnamefont {M.}~\bibnamefont
  {Tylutki}}\ and\ \bibinfo {author} {\bibfnamefont {P.}~\bibnamefont
  {T\"orm\"a}},\ }\href {\doibase 10.1103/PhysRevB.98.094513} {\bibfield
  {journal} {\bibinfo  {journal} {Phys. Rev. B}\ }\textbf {\bibinfo {volume}
  {98}},\ \bibinfo {pages} {094513} (\bibinfo {year} {2018})}\BibitemShut
  {NoStop}%
\bibitem [{\citenamefont {Di~Liberto}\ \emph {et~al.}(2019)\citenamefont
  {Di~Liberto}, \citenamefont {Mukherjee},\ and\ \citenamefont
  {Goldman}}]{PhysRevA.100.043829}%
  \BibitemOpen
  \bibfield  {author} {\bibinfo {author} {\bibfnamefont {M.}~\bibnamefont
  {Di~Liberto}}, \bibinfo {author} {\bibfnamefont {S.}~\bibnamefont
  {Mukherjee}}, \ and\ \bibinfo {author} {\bibfnamefont {N.}~\bibnamefont
  {Goldman}},\ }\href {\doibase 10.1103/PhysRevA.100.043829} {\bibfield
  {journal} {\bibinfo  {journal} {Phys. Rev. A}\ }\textbf {\bibinfo {volume}
  {100}},\ \bibinfo {pages} {043829} (\bibinfo {year} {2019})}\BibitemShut
  {NoStop}%
\bibitem [{\citenamefont {Kuno}(2020)}]{PhysRevB.101.184112}%
  \BibitemOpen
  \bibfield  {author} {\bibinfo {author} {\bibfnamefont {Y.}~\bibnamefont
  {Kuno}},\ }\href {\doibase 10.1103/PhysRevB.101.184112} {\bibfield  {journal}
  {\bibinfo  {journal} {Phys. Rev. B}\ }\textbf {\bibinfo {volume} {101}},\
  \bibinfo {pages} {184112} (\bibinfo {year} {2020})}\BibitemShut {NoStop}%
\bibitem [{\citenamefont {Pelegr{\'\i}}\ \emph {et~al.}(2020)\citenamefont
  {Pelegr{\'\i}}, \citenamefont {Marques}, \citenamefont {Ahufinger},
  \citenamefont {Mompart},\ and\ \citenamefont {Dias}}]{GPel}%
  \BibitemOpen
  \bibfield  {author} {\bibinfo {author} {\bibfnamefont {G.}~\bibnamefont
  {Pelegr{\'\i}}}, \bibinfo {author} {\bibfnamefont {A.~M.}\ \bibnamefont
  {Marques}}, \bibinfo {author} {\bibfnamefont {V.}~\bibnamefont {Ahufinger}},
  \bibinfo {author} {\bibfnamefont {J.}~\bibnamefont {Mompart}}, \ and\
  \bibinfo {author} {\bibfnamefont {R.~G.}\ \bibnamefont {Dias}},\ }\href@noop
  {} {\bibfield  {journal} {\bibinfo  {journal} {arXiv:2005.10810}\ } (\bibinfo
  {year} {2020})}\BibitemShut {NoStop}%
\bibitem [{\citenamefont {Kuno}\ \emph {et~al.}(2020)\citenamefont {Kuno},
  \citenamefont {Orito},\ and\ \citenamefont {Ichinose}}]{Kuno:2020aa}%
  \BibitemOpen
  \bibfield  {author} {\bibinfo {author} {\bibfnamefont {Y.}~\bibnamefont
  {Kuno}}, \bibinfo {author} {\bibfnamefont {T.}~\bibnamefont {Orito}}, \ and\
  \bibinfo {author} {\bibfnamefont {I.}~\bibnamefont {Ichinose}},\ }\href
  {\doibase 10.1088/1367-2630/ab6352} {\bibfield  {journal} {\bibinfo
  {journal} {New Journal of Physics}\ }\textbf {\bibinfo {volume} {22}},\
  \bibinfo {pages} {013032} (\bibinfo {year} {2020})}\BibitemShut {NoStop}%
\bibitem [{\citenamefont {Salerno}\ \emph {et~al.}(2020)\citenamefont
  {Salerno}, \citenamefont {Palumbo}, \citenamefont {Goldman},\ and\
  \citenamefont {Di~Liberto}}]{PhysRevResearch.2.013348}%
  \BibitemOpen
  \bibfield  {author} {\bibinfo {author} {\bibfnamefont {G.}~\bibnamefont
  {Salerno}}, \bibinfo {author} {\bibfnamefont {G.}~\bibnamefont {Palumbo}},
  \bibinfo {author} {\bibfnamefont {N.}~\bibnamefont {Goldman}}, \ and\
  \bibinfo {author} {\bibfnamefont {M.}~\bibnamefont {Di~Liberto}},\ }\href
  {\doibase 10.1103/PhysRevResearch.2.013348} {\bibfield  {journal} {\bibinfo
  {journal} {Phys. Rev. Research}\ }\textbf {\bibinfo {volume} {2}},\ \bibinfo
  {pages} {013348} (\bibinfo {year} {2020})}\BibitemShut {NoStop}%
\bibitem [{\citenamefont {Mondaini}\ \emph {et~al.}(2018)\citenamefont
  {Mondaini}, \citenamefont {Batrouni},\ and\ \citenamefont
  {Gr\'emaud}}]{PhysRevB.98.155142}%
  \BibitemOpen
  \bibfield  {author} {\bibinfo {author} {\bibfnamefont {R.}~\bibnamefont
  {Mondaini}}, \bibinfo {author} {\bibfnamefont {G.~G.}\ \bibnamefont
  {Batrouni}}, \ and\ \bibinfo {author} {\bibfnamefont {B.}~\bibnamefont
  {Gr\'emaud}},\ }\href {\doibase 10.1103/PhysRevB.98.155142} {\bibfield
  {journal} {\bibinfo  {journal} {Phys. Rev. B}\ }\textbf {\bibinfo {volume}
  {98}},\ \bibinfo {pages} {155142} (\bibinfo {year} {2018})}\BibitemShut
  {NoStop}%
\bibitem [{\citenamefont {Zurita}\ \emph {et~al.}(2020)\citenamefont {Zurita},
  \citenamefont {Creffield},\ and\ \citenamefont {Platero}}]{Zurita:2020aa}%
  \BibitemOpen
  \bibfield  {author} {\bibinfo {author} {\bibfnamefont {J.}~\bibnamefont
  {Zurita}}, \bibinfo {author} {\bibfnamefont {C.~E.}\ \bibnamefont
  {Creffield}}, \ and\ \bibinfo {author} {\bibfnamefont {G.}~\bibnamefont
  {Platero}},\ }\bibfield  {booktitle} {\emph {\bibinfo {booktitle} {Advanced
  Quantum Technologies}},\ }\href {\doibase 10.1002/qute.201900105} {\bibfield
  {journal} {\bibinfo  {journal} {Advanced Quantum Technologies}\ }\textbf
  {\bibinfo {volume} {3}},\ \bibinfo {pages} {1900105} (\bibinfo {year}
  {2020})}\BibitemShut {NoStop}%
\bibitem [{\citenamefont {Cartwright}\ \emph {et~al.}(2018)\citenamefont
  {Cartwright}, \citenamefont {De~Chiara},\ and\ \citenamefont
  {Rizzi}}]{PhysRevB.98.184508}%
  \BibitemOpen
  \bibfield  {author} {\bibinfo {author} {\bibfnamefont {C.}~\bibnamefont
  {Cartwright}}, \bibinfo {author} {\bibfnamefont {G.}~\bibnamefont
  {De~Chiara}}, \ and\ \bibinfo {author} {\bibfnamefont {M.}~\bibnamefont
  {Rizzi}},\ }\href {\doibase 10.1103/PhysRevB.98.184508} {\bibfield  {journal}
  {\bibinfo  {journal} {Phys. Rev. B}\ }\textbf {\bibinfo {volume} {98}},\
  \bibinfo {pages} {184508} (\bibinfo {year} {2018})}\BibitemShut {NoStop}%
\bibitem [{\citenamefont {Winkler}\ \emph {et~al.}(2006)\citenamefont
  {Winkler}, \citenamefont {Thalhammer}, \citenamefont {Lang}, \citenamefont
  {Grimm}, \citenamefont {Hecker~Denschlag}, \citenamefont {Daley},
  \citenamefont {Kantian}, \citenamefont {B{\"u}chler},\ and\ \citenamefont
  {Zoller}}]{Winkler:2006aa}%
  \BibitemOpen
  \bibfield  {author} {\bibinfo {author} {\bibfnamefont {K.}~\bibnamefont
  {Winkler}}, \bibinfo {author} {\bibfnamefont {G.}~\bibnamefont {Thalhammer}},
  \bibinfo {author} {\bibfnamefont {F.}~\bibnamefont {Lang}}, \bibinfo {author}
  {\bibfnamefont {R.}~\bibnamefont {Grimm}}, \bibinfo {author} {\bibfnamefont
  {J.}~\bibnamefont {Hecker~Denschlag}}, \bibinfo {author} {\bibfnamefont
  {A.~J.}\ \bibnamefont {Daley}}, \bibinfo {author} {\bibfnamefont
  {A.}~\bibnamefont {Kantian}}, \bibinfo {author} {\bibfnamefont {H.~P.}\
  \bibnamefont {B{\"u}chler}}, \ and\ \bibinfo {author} {\bibfnamefont
  {P.}~\bibnamefont {Zoller}},\ }\href {\doibase 10.1038/nature04918}
  {\bibfield  {journal} {\bibinfo  {journal} {Nature}\ }\textbf {\bibinfo
  {volume} {441}},\ \bibinfo {pages} {853} (\bibinfo {year}
  {2006})}\BibitemShut {NoStop}%
\bibitem [{\citenamefont {Deuchert}\ \emph {et~al.}(2012)\citenamefont
  {Deuchert}, \citenamefont {Sakmann}, \citenamefont {Streltsov}, \citenamefont
  {Alon},\ and\ \citenamefont {Cederbaum}}]{PhysRevA.86.013618}%
  \BibitemOpen
  \bibfield  {author} {\bibinfo {author} {\bibfnamefont {A.}~\bibnamefont
  {Deuchert}}, \bibinfo {author} {\bibfnamefont {K.}~\bibnamefont {Sakmann}},
  \bibinfo {author} {\bibfnamefont {A.~I.}\ \bibnamefont {Streltsov}}, \bibinfo
  {author} {\bibfnamefont {O.~E.}\ \bibnamefont {Alon}}, \ and\ \bibinfo
  {author} {\bibfnamefont {L.~S.}\ \bibnamefont {Cederbaum}},\ }\href {\doibase
  10.1103/PhysRevA.86.013618} {\bibfield  {journal} {\bibinfo  {journal} {Phys.
  Rev. A}\ }\textbf {\bibinfo {volume} {86}},\ \bibinfo {pages} {013618}
  (\bibinfo {year} {2012})}\BibitemShut {NoStop}%
\bibitem [{\citenamefont {Lin}\ \emph {et~al.}(2020)\citenamefont {Lin},
  \citenamefont {Ke},\ and\ \citenamefont {Lee}}]{PhysRevA.101.023620}%
  \BibitemOpen
  \bibfield  {author} {\bibinfo {author} {\bibfnamefont {L.}~\bibnamefont
  {Lin}}, \bibinfo {author} {\bibfnamefont {Y.}~\bibnamefont {Ke}}, \ and\
  \bibinfo {author} {\bibfnamefont {C.}~\bibnamefont {Lee}},\ }\href {\doibase
  10.1103/PhysRevA.101.023620} {\bibfield  {journal} {\bibinfo  {journal}
  {Phys. Rev. A}\ }\textbf {\bibinfo {volume} {101}},\ \bibinfo {pages}
  {023620} (\bibinfo {year} {2020})}\BibitemShut {NoStop}%
\bibitem [{\citenamefont {Ke}\ \emph {et~al.}(2017)\citenamefont {Ke},
  \citenamefont {Qin}, \citenamefont {Kivshar},\ and\ \citenamefont
  {Lee}}]{PhysRevA.95.063630}%
  \BibitemOpen
  \bibfield  {author} {\bibinfo {author} {\bibfnamefont {Y.}~\bibnamefont
  {Ke}}, \bibinfo {author} {\bibfnamefont {X.}~\bibnamefont {Qin}}, \bibinfo
  {author} {\bibfnamefont {Y.~S.}\ \bibnamefont {Kivshar}}, \ and\ \bibinfo
  {author} {\bibfnamefont {C.}~\bibnamefont {Lee}},\ }\href {\doibase
  10.1103/PhysRevA.95.063630} {\bibfield  {journal} {\bibinfo  {journal} {Phys.
  Rev. A}\ }\textbf {\bibinfo {volume} {95}},\ \bibinfo {pages} {063630}
  (\bibinfo {year} {2017})}\BibitemShut {NoStop}%
\bibitem [{\citenamefont {Qin}\ \emph {et~al.}(2017)\citenamefont {Qin},
  \citenamefont {Mei}, \citenamefont {Ke}, \citenamefont {Zhang},\ and\
  \citenamefont {Lee}}]{PhysRevB.96.195134}%
  \BibitemOpen
  \bibfield  {author} {\bibinfo {author} {\bibfnamefont {X.}~\bibnamefont
  {Qin}}, \bibinfo {author} {\bibfnamefont {F.}~\bibnamefont {Mei}}, \bibinfo
  {author} {\bibfnamefont {Y.}~\bibnamefont {Ke}}, \bibinfo {author}
  {\bibfnamefont {L.}~\bibnamefont {Zhang}}, \ and\ \bibinfo {author}
  {\bibfnamefont {C.}~\bibnamefont {Lee}},\ }\href {\doibase
  10.1103/PhysRevB.96.195134} {\bibfield  {journal} {\bibinfo  {journal} {Phys.
  Rev. B}\ }\textbf {\bibinfo {volume} {96}},\ \bibinfo {pages} {195134}
  (\bibinfo {year} {2017})}\BibitemShut {NoStop}%
\bibitem [{\citenamefont {Qin}\ \emph {et~al.}(2018)\citenamefont {Qin},
  \citenamefont {Mei}, \citenamefont {Ke}, \citenamefont {Zhang},\ and\
  \citenamefont {Lee}}]{Qin:2018aa}%
  \BibitemOpen
  \bibfield  {author} {\bibinfo {author} {\bibfnamefont {X.}~\bibnamefont
  {Qin}}, \bibinfo {author} {\bibfnamefont {F.}~\bibnamefont {Mei}}, \bibinfo
  {author} {\bibfnamefont {Y.}~\bibnamefont {Ke}}, \bibinfo {author}
  {\bibfnamefont {L.}~\bibnamefont {Zhang}}, \ and\ \bibinfo {author}
  {\bibfnamefont {C.}~\bibnamefont {Lee}},\ }\bibfield  {booktitle} {\emph
  {\bibinfo {booktitle} {New Journal of Physics}},\ }\href {\doibase
  10.1088/1367-2630/aa9556} {\ \textbf {\bibinfo {volume} {20}},\ \bibinfo
  {pages} {013003} (\bibinfo {year} {2018})}\BibitemShut {NoStop}%
\bibitem [{\citenamefont {Di~Liberto}\ \emph
  {et~al.}(2016{\natexlab{b}})\citenamefont {Di~Liberto}, \citenamefont
  {Recati}, \citenamefont {Carusotto},\ and\ \citenamefont
  {Menotti}}]{PhysRevA.94.062704}%
  \BibitemOpen
  \bibfield  {author} {\bibinfo {author} {\bibfnamefont {M.}~\bibnamefont
  {Di~Liberto}}, \bibinfo {author} {\bibfnamefont {A.}~\bibnamefont {Recati}},
  \bibinfo {author} {\bibfnamefont {I.}~\bibnamefont {Carusotto}}, \ and\
  \bibinfo {author} {\bibfnamefont {C.}~\bibnamefont {Menotti}},\ }\href
  {\doibase 10.1103/PhysRevA.94.062704} {\bibfield  {journal} {\bibinfo
  {journal} {Phys. Rev. A}\ }\textbf {\bibinfo {volume} {94}},\ \bibinfo
  {pages} {062704} (\bibinfo {year} {2016}{\natexlab{b}})}\BibitemShut
  {NoStop}%
\bibitem [{\citenamefont {Marques}\ and\ \citenamefont
  {Dias}(2018)}]{Marques_2018}%
  \BibitemOpen
  \bibfield  {author} {\bibinfo {author} {\bibfnamefont {A.~M.}\ \bibnamefont
  {Marques}}\ and\ \bibinfo {author} {\bibfnamefont {R.~G.}\ \bibnamefont
  {Dias}},\ }\href {\doibase 10.1088/1361-648x/aacd7c} {\bibfield  {journal}
  {\bibinfo  {journal} {Journal of Physics: Condensed Matter}\ }\textbf
  {\bibinfo {volume} {30}},\ \bibinfo {pages} {305601} (\bibinfo {year}
  {2018})}\BibitemShut {NoStop}%
\bibitem [{\citenamefont {Julku}\ \emph {et~al.}(2016)\citenamefont {Julku},
  \citenamefont {Peotta}, \citenamefont {Vanhala}, \citenamefont {Kim},\ and\
  \citenamefont {T\"orm\"a}}]{PhysRevLett.117.045303}%
  \BibitemOpen
  \bibfield  {author} {\bibinfo {author} {\bibfnamefont {A.}~\bibnamefont
  {Julku}}, \bibinfo {author} {\bibfnamefont {S.}~\bibnamefont {Peotta}},
  \bibinfo {author} {\bibfnamefont {T.~I.}\ \bibnamefont {Vanhala}}, \bibinfo
  {author} {\bibfnamefont {D.-H.}\ \bibnamefont {Kim}}, \ and\ \bibinfo
  {author} {\bibfnamefont {P.}~\bibnamefont {T\"orm\"a}},\ }\href {\doibase
  10.1103/PhysRevLett.117.045303} {\bibfield  {journal} {\bibinfo  {journal}
  {Phys. Rev. Lett.}\ }\textbf {\bibinfo {volume} {117}},\ \bibinfo {pages}
  {045303} (\bibinfo {year} {2016})}\BibitemShut {NoStop}%
\bibitem [{\citenamefont {Mishra}\ \emph
  {et~al.}(2015{\natexlab{a}})\citenamefont {Mishra}, \citenamefont
  {Greschner},\ and\ \citenamefont {Santos}}]{PhysRevB.92.195149}%
  \BibitemOpen
  \bibfield  {author} {\bibinfo {author} {\bibfnamefont {T.}~\bibnamefont
  {Mishra}}, \bibinfo {author} {\bibfnamefont {S.}~\bibnamefont {Greschner}}, \
  and\ \bibinfo {author} {\bibfnamefont {L.}~\bibnamefont {Santos}},\ }\href
  {\doibase 10.1103/PhysRevB.92.195149} {\bibfield  {journal} {\bibinfo
  {journal} {Phys. Rev. B}\ }\textbf {\bibinfo {volume} {92}},\ \bibinfo
  {pages} {195149} (\bibinfo {year} {2015}{\natexlab{a}})}\BibitemShut
  {NoStop}%
\bibitem [{\citenamefont {Mishra}\ \emph
  {et~al.}(2015{\natexlab{b}})\citenamefont {Mishra}, \citenamefont
  {Greschner},\ and\ \citenamefont {Santos}}]{PhysRevA.91.043614}%
  \BibitemOpen
  \bibfield  {author} {\bibinfo {author} {\bibfnamefont {T.}~\bibnamefont
  {Mishra}}, \bibinfo {author} {\bibfnamefont {S.}~\bibnamefont {Greschner}}, \
  and\ \bibinfo {author} {\bibfnamefont {L.}~\bibnamefont {Santos}},\ }\href
  {\doibase 10.1103/PhysRevA.91.043614} {\bibfield  {journal} {\bibinfo
  {journal} {Phys. Rev. A}\ }\textbf {\bibinfo {volume} {91}},\ \bibinfo
  {pages} {043614} (\bibinfo {year} {2015}{\natexlab{b}})}\BibitemShut
  {NoStop}%
\bibitem [{\citenamefont {Scarola}\ and\ \citenamefont
  {Das~Sarma}(2005)}]{PhysRevLett.95.033003}%
  \BibitemOpen
  \bibfield  {author} {\bibinfo {author} {\bibfnamefont {V.~W.}\ \bibnamefont
  {Scarola}}\ and\ \bibinfo {author} {\bibfnamefont {S.}~\bibnamefont
  {Das~Sarma}},\ }\href {\doibase 10.1103/PhysRevLett.95.033003} {\bibfield
  {journal} {\bibinfo  {journal} {Phys. Rev. Lett.}\ }\textbf {\bibinfo
  {volume} {95}},\ \bibinfo {pages} {033003} (\bibinfo {year}
  {2005})}\BibitemShut {NoStop}%
\bibitem [{\citenamefont {Scarola}\ \emph {et~al.}(2006)\citenamefont
  {Scarola}, \citenamefont {Demler},\ and\ \citenamefont
  {Das~Sarma}}]{PhysRevA.73.051601}%
  \BibitemOpen
  \bibfield  {author} {\bibinfo {author} {\bibfnamefont {V.~W.}\ \bibnamefont
  {Scarola}}, \bibinfo {author} {\bibfnamefont {E.}~\bibnamefont {Demler}}, \
  and\ \bibinfo {author} {\bibfnamefont {S.}~\bibnamefont {Das~Sarma}},\ }\href
  {\doibase 10.1103/PhysRevA.73.051601} {\bibfield  {journal} {\bibinfo
  {journal} {Phys. Rev. A}\ }\textbf {\bibinfo {volume} {73}},\ \bibinfo
  {pages} {051601} (\bibinfo {year} {2006})}\BibitemShut {NoStop}%
\bibitem [{\citenamefont {Iskin}(2011)}]{PhysRevA.83.051606}%
  \BibitemOpen
  \bibfield  {author} {\bibinfo {author} {\bibfnamefont {M.}~\bibnamefont
  {Iskin}},\ }\href {\doibase 10.1103/PhysRevA.83.051606} {\bibfield  {journal}
  {\bibinfo  {journal} {Phys. Rev. A}\ }\textbf {\bibinfo {volume} {83}},\
  \bibinfo {pages} {051606} (\bibinfo {year} {2011})}\BibitemShut {NoStop}%
\bibitem [{\citenamefont {Wang}\ and\ \citenamefont {Liang}(2012)}]{Wang_2012}%
  \BibitemOpen
  \bibfield  {author} {\bibinfo {author} {\bibfnamefont {Y.-M.}\ \bibnamefont
  {Wang}}\ and\ \bibinfo {author} {\bibfnamefont {J.-Q.}\ \bibnamefont
  {Liang}},\ }\href {\doibase 10.1088/1674-1056/21/6/060305} {\bibfield
  {journal} {\bibinfo  {journal} {Chinese Physics B}\ }\textbf {\bibinfo
  {volume} {21}},\ \bibinfo {pages} {060305} (\bibinfo {year}
  {2012})}\BibitemShut {NoStop}%
\bibitem [{\citenamefont {Mathey}\ \emph {et~al.}(2009)\citenamefont {Mathey},
  \citenamefont {Danshita},\ and\ \citenamefont {Clark}}]{PhysRevA.79.011602}%
  \BibitemOpen
  \bibfield  {author} {\bibinfo {author} {\bibfnamefont {L.}~\bibnamefont
  {Mathey}}, \bibinfo {author} {\bibfnamefont {I.}~\bibnamefont {Danshita}}, \
  and\ \bibinfo {author} {\bibfnamefont {C.~W.}\ \bibnamefont {Clark}},\ }\href
  {\doibase 10.1103/PhysRevA.79.011602} {\bibfield  {journal} {\bibinfo
  {journal} {Phys. Rev. A}\ }\textbf {\bibinfo {volume} {79}},\ \bibinfo
  {pages} {011602} (\bibinfo {year} {2009})}\BibitemShut {NoStop}%
\bibitem [{\citenamefont {Danshita}\ and\ \citenamefont {S\'a~de
  Melo}(2009)}]{PhysRevLett.103.225301}%
  \BibitemOpen
  \bibfield  {author} {\bibinfo {author} {\bibfnamefont {I.}~\bibnamefont
  {Danshita}}\ and\ \bibinfo {author} {\bibfnamefont {C.~A.~R.}\ \bibnamefont
  {S\'a~de Melo}},\ }\href {\doibase 10.1103/PhysRevLett.103.225301} {\bibfield
   {journal} {\bibinfo  {journal} {Phys. Rev. Lett.}\ }\textbf {\bibinfo
  {volume} {103}},\ \bibinfo {pages} {225301} (\bibinfo {year}
  {2009})}\BibitemShut {NoStop}%
\bibitem [{\citenamefont {Diehl}\ \emph {et~al.}(2010)\citenamefont {Diehl},
  \citenamefont {Baranov}, \citenamefont {Daley},\ and\ \citenamefont
  {Zoller}}]{PhysRevLett.104.165301}%
  \BibitemOpen
  \bibfield  {author} {\bibinfo {author} {\bibfnamefont {S.}~\bibnamefont
  {Diehl}}, \bibinfo {author} {\bibfnamefont {M.}~\bibnamefont {Baranov}},
  \bibinfo {author} {\bibfnamefont {A.~J.}\ \bibnamefont {Daley}}, \ and\
  \bibinfo {author} {\bibfnamefont {P.}~\bibnamefont {Zoller}},\ }\href
  {\doibase 10.1103/PhysRevLett.104.165301} {\bibfield  {journal} {\bibinfo
  {journal} {Phys. Rev. Lett.}\ }\textbf {\bibinfo {volume} {104}},\ \bibinfo
  {pages} {165301} (\bibinfo {year} {2010})}\BibitemShut {NoStop}%
\bibitem [{\citenamefont {Trefzger}\ \emph {et~al.}(2009)\citenamefont
  {Trefzger}, \citenamefont {Menotti},\ and\ \citenamefont
  {Lewenstein}}]{PhysRevLett.103.035304}%
  \BibitemOpen
  \bibfield  {author} {\bibinfo {author} {\bibfnamefont {C.}~\bibnamefont
  {Trefzger}}, \bibinfo {author} {\bibfnamefont {C.}~\bibnamefont {Menotti}}, \
  and\ \bibinfo {author} {\bibfnamefont {M.}~\bibnamefont {Lewenstein}},\
  }\href {\doibase 10.1103/PhysRevLett.103.035304} {\bibfield  {journal}
  {\bibinfo  {journal} {Phys. Rev. Lett.}\ }\textbf {\bibinfo {volume} {103}},\
  \bibinfo {pages} {035304} (\bibinfo {year} {2009})}\BibitemShut {NoStop}%
\bibitem [{\citenamefont {Hu}\ \emph {et~al.}(2009)\citenamefont {Hu},
  \citenamefont {Mathey}, \citenamefont {Danshita}, \citenamefont {Tiesinga},
  \citenamefont {Williams},\ and\ \citenamefont {Clark}}]{PhysRevA.80.023619}%
  \BibitemOpen
  \bibfield  {author} {\bibinfo {author} {\bibfnamefont {A.}~\bibnamefont
  {Hu}}, \bibinfo {author} {\bibfnamefont {L.}~\bibnamefont {Mathey}}, \bibinfo
  {author} {\bibfnamefont {I.}~\bibnamefont {Danshita}}, \bibinfo {author}
  {\bibfnamefont {E.}~\bibnamefont {Tiesinga}}, \bibinfo {author}
  {\bibfnamefont {C.~J.}\ \bibnamefont {Williams}}, \ and\ \bibinfo {author}
  {\bibfnamefont {C.~W.}\ \bibnamefont {Clark}},\ }\href {\doibase
  10.1103/PhysRevA.80.023619} {\bibfield  {journal} {\bibinfo  {journal} {Phys.
  Rev. A}\ }\textbf {\bibinfo {volume} {80}},\ \bibinfo {pages} {023619}
  (\bibinfo {year} {2009})}\BibitemShut {NoStop}%
\bibitem [{\citenamefont {Richerme}\ \emph {et~al.}(2013)\citenamefont
  {Richerme}, \citenamefont {Senko}, \citenamefont {Smith}, \citenamefont
  {Lee}, \citenamefont {Korenblit},\ and\ \citenamefont
  {Monroe}}]{PhysRevA.88.012334}%
  \BibitemOpen
  \bibfield  {author} {\bibinfo {author} {\bibfnamefont {P.}~\bibnamefont
  {Richerme}}, \bibinfo {author} {\bibfnamefont {C.}~\bibnamefont {Senko}},
  \bibinfo {author} {\bibfnamefont {J.}~\bibnamefont {Smith}}, \bibinfo
  {author} {\bibfnamefont {A.}~\bibnamefont {Lee}}, \bibinfo {author}
  {\bibfnamefont {S.}~\bibnamefont {Korenblit}}, \ and\ \bibinfo {author}
  {\bibfnamefont {C.}~\bibnamefont {Monroe}},\ }\href {\doibase
  10.1103/PhysRevA.88.012334} {\bibfield  {journal} {\bibinfo  {journal} {Phys.
  Rev. A}\ }\textbf {\bibinfo {volume} {88}},\ \bibinfo {pages} {012334}
  (\bibinfo {year} {2013})}\BibitemShut {NoStop}%
\bibitem [{\citenamefont {Simon}\ \emph {et~al.}(2011)\citenamefont {Simon},
  \citenamefont {Bakr}, \citenamefont {Ma}, \citenamefont {Tai}, \citenamefont
  {Preiss},\ and\ \citenamefont {Greiner}}]{Simon:2011aa}%
  \BibitemOpen
  \bibfield  {author} {\bibinfo {author} {\bibfnamefont {J.}~\bibnamefont
  {Simon}}, \bibinfo {author} {\bibfnamefont {W.~S.}\ \bibnamefont {Bakr}},
  \bibinfo {author} {\bibfnamefont {R.}~\bibnamefont {Ma}}, \bibinfo {author}
  {\bibfnamefont {M.~E.}\ \bibnamefont {Tai}}, \bibinfo {author} {\bibfnamefont
  {P.~M.}\ \bibnamefont {Preiss}}, \ and\ \bibinfo {author} {\bibfnamefont
  {M.}~\bibnamefont {Greiner}},\ }\href {\doibase 10.1038/nature09994}
  {\bibfield  {journal} {\bibinfo  {journal} {Nature}\ }\textbf {\bibinfo
  {volume} {472}},\ \bibinfo {pages} {307} (\bibinfo {year}
  {2011})}\BibitemShut {NoStop}%
\bibitem [{\citenamefont {Mugel}\ \emph {et~al.}(2016)\citenamefont {Mugel},
  \citenamefont {Celi}, \citenamefont {Massignan}, \citenamefont {Asb\'oth},
  \citenamefont {Lewenstein},\ and\ \citenamefont {Lobo}}]{PhysRevA.94.023631}%
  \BibitemOpen
  \bibfield  {author} {\bibinfo {author} {\bibfnamefont {S.}~\bibnamefont
  {Mugel}}, \bibinfo {author} {\bibfnamefont {A.}~\bibnamefont {Celi}},
  \bibinfo {author} {\bibfnamefont {P.}~\bibnamefont {Massignan}}, \bibinfo
  {author} {\bibfnamefont {J.~K.}\ \bibnamefont {Asb\'oth}}, \bibinfo {author}
  {\bibfnamefont {M.}~\bibnamefont {Lewenstein}}, \ and\ \bibinfo {author}
  {\bibfnamefont {C.}~\bibnamefont {Lobo}},\ }\href {\doibase
  10.1103/PhysRevA.94.023631} {\bibfield  {journal} {\bibinfo  {journal} {Phys.
  Rev. A}\ }\textbf {\bibinfo {volume} {94}},\ \bibinfo {pages} {023631}
  (\bibinfo {year} {2016})}\BibitemShut {NoStop}%
\bibitem [{\citenamefont {Kang}\ \emph {et~al.}(2020)\citenamefont {Kang},
  \citenamefont {Han},\ and\ \citenamefont {Shin}}]{Kang:2020aa}%
  \BibitemOpen
  \bibfield  {author} {\bibinfo {author} {\bibfnamefont {J.~H.}\ \bibnamefont
  {Kang}}, \bibinfo {author} {\bibfnamefont {J.~H.}\ \bibnamefont {Han}}, \
  and\ \bibinfo {author} {\bibfnamefont {Y.}~\bibnamefont {Shin}},\ }\href
  {\doibase 10.1088/1367-2630/ab61d7} {\bibfield  {journal} {\bibinfo
  {journal} {New Journal of Physics}\ }\textbf {\bibinfo {volume} {22}},\
  \bibinfo {pages} {013023} (\bibinfo {year} {2020})}\BibitemShut {NoStop}%
\bibitem [{\citenamefont {Livi}\ \emph {et~al.}(2016)\citenamefont {Livi},
  \citenamefont {Cappellini}, \citenamefont {Diem}, \citenamefont {Franchi},
  \citenamefont {Clivati}, \citenamefont {Frittelli}, \citenamefont {Levi},
  \citenamefont {Calonico}, \citenamefont {Catani}, \citenamefont {Inguscio},\
  and\ \citenamefont {Fallani}}]{PhysRevLett.117.220401}%
  \BibitemOpen
  \bibfield  {author} {\bibinfo {author} {\bibfnamefont {L.~F.}\ \bibnamefont
  {Livi}}, \bibinfo {author} {\bibfnamefont {G.}~\bibnamefont {Cappellini}},
  \bibinfo {author} {\bibfnamefont {M.}~\bibnamefont {Diem}}, \bibinfo {author}
  {\bibfnamefont {L.}~\bibnamefont {Franchi}}, \bibinfo {author} {\bibfnamefont
  {C.}~\bibnamefont {Clivati}}, \bibinfo {author} {\bibfnamefont
  {M.}~\bibnamefont {Frittelli}}, \bibinfo {author} {\bibfnamefont
  {F.}~\bibnamefont {Levi}}, \bibinfo {author} {\bibfnamefont {D.}~\bibnamefont
  {Calonico}}, \bibinfo {author} {\bibfnamefont {J.}~\bibnamefont {Catani}},
  \bibinfo {author} {\bibfnamefont {M.}~\bibnamefont {Inguscio}}, \ and\
  \bibinfo {author} {\bibfnamefont {L.}~\bibnamefont {Fallani}},\ }\href
  {\doibase 10.1103/PhysRevLett.117.220401} {\bibfield  {journal} {\bibinfo
  {journal} {Phys. Rev. Lett.}\ }\textbf {\bibinfo {volume} {117}},\ \bibinfo
  {pages} {220401} (\bibinfo {year} {2016})}\BibitemShut {NoStop}%
\bibitem [{\citenamefont {Boada}\ \emph {et~al.}(2012)\citenamefont {Boada},
  \citenamefont {Celi}, \citenamefont {Latorre},\ and\ \citenamefont
  {Lewenstein}}]{PhysRevLett.108.133001}%
  \BibitemOpen
  \bibfield  {author} {\bibinfo {author} {\bibfnamefont {O.}~\bibnamefont
  {Boada}}, \bibinfo {author} {\bibfnamefont {A.}~\bibnamefont {Celi}},
  \bibinfo {author} {\bibfnamefont {J.~I.}\ \bibnamefont {Latorre}}, \ and\
  \bibinfo {author} {\bibfnamefont {M.}~\bibnamefont {Lewenstein}},\ }\href
  {\doibase 10.1103/PhysRevLett.108.133001} {\bibfield  {journal} {\bibinfo
  {journal} {Phys. Rev. Lett.}\ }\textbf {\bibinfo {volume} {108}},\ \bibinfo
  {pages} {133001} (\bibinfo {year} {2012})}\BibitemShut {NoStop}%
\bibitem [{\citenamefont {Celi}\ \emph {et~al.}(2014)\citenamefont {Celi},
  \citenamefont {Massignan}, \citenamefont {Ruseckas}, \citenamefont {Goldman},
  \citenamefont {Spielman}, \citenamefont {Juzeli\ifmmode~\bar{u}\else
  \={u}\fi{}nas},\ and\ \citenamefont {Lewenstein}}]{PhysRevLett.112.043001}%
  \BibitemOpen
  \bibfield  {author} {\bibinfo {author} {\bibfnamefont {A.}~\bibnamefont
  {Celi}}, \bibinfo {author} {\bibfnamefont {P.}~\bibnamefont {Massignan}},
  \bibinfo {author} {\bibfnamefont {J.}~\bibnamefont {Ruseckas}}, \bibinfo
  {author} {\bibfnamefont {N.}~\bibnamefont {Goldman}}, \bibinfo {author}
  {\bibfnamefont {I.~B.}\ \bibnamefont {Spielman}}, \bibinfo {author}
  {\bibfnamefont {G.}~\bibnamefont {Juzeli\ifmmode~\bar{u}\else
  \={u}\fi{}nas}}, \ and\ \bibinfo {author} {\bibfnamefont {M.}~\bibnamefont
  {Lewenstein}},\ }\href {\doibase 10.1103/PhysRevLett.112.043001} {\bibfield
  {journal} {\bibinfo  {journal} {Phys. Rev. Lett.}\ }\textbf {\bibinfo
  {volume} {112}},\ \bibinfo {pages} {043001} (\bibinfo {year}
  {2014})}\BibitemShut {NoStop}%
\bibitem [{\citenamefont {Mancini}\ \emph {et~al.}(2015)\citenamefont
  {Mancini}, \citenamefont {Pagano}, \citenamefont {Cappellini}, \citenamefont
  {Livi}, \citenamefont {Rider}, \citenamefont {Catani}, \citenamefont {Sias},
  \citenamefont {Zoller}, \citenamefont {Inguscio}, \citenamefont {Dalmonte},\
  and\ \citenamefont {Fallani}}]{Mancini1510}%
  \BibitemOpen
  \bibfield  {author} {\bibinfo {author} {\bibfnamefont {M.}~\bibnamefont
  {Mancini}}, \bibinfo {author} {\bibfnamefont {G.}~\bibnamefont {Pagano}},
  \bibinfo {author} {\bibfnamefont {G.}~\bibnamefont {Cappellini}}, \bibinfo
  {author} {\bibfnamefont {L.}~\bibnamefont {Livi}}, \bibinfo {author}
  {\bibfnamefont {M.}~\bibnamefont {Rider}}, \bibinfo {author} {\bibfnamefont
  {J.}~\bibnamefont {Catani}}, \bibinfo {author} {\bibfnamefont
  {C.}~\bibnamefont {Sias}}, \bibinfo {author} {\bibfnamefont {P.}~\bibnamefont
  {Zoller}}, \bibinfo {author} {\bibfnamefont {M.}~\bibnamefont {Inguscio}},
  \bibinfo {author} {\bibfnamefont {M.}~\bibnamefont {Dalmonte}}, \ and\
  \bibinfo {author} {\bibfnamefont {L.}~\bibnamefont {Fallani}},\ }\href
  {\doibase 10.1126/science.aaa8736} {\bibfield  {journal} {\bibinfo  {journal}
  {Science}\ }\textbf {\bibinfo {volume} {349}},\ \bibinfo {pages} {1510}
  (\bibinfo {year} {2015})}\BibitemShut {NoStop}%
\bibitem [{\citenamefont {Stuhl}\ \emph {et~al.}(2015)\citenamefont {Stuhl},
  \citenamefont {Lu}, \citenamefont {Aycock}, \citenamefont {Genkina},\ and\
  \citenamefont {Spielman}}]{Stuhl1514}%
  \BibitemOpen
  \bibfield  {author} {\bibinfo {author} {\bibfnamefont {B.~K.}\ \bibnamefont
  {Stuhl}}, \bibinfo {author} {\bibfnamefont {H.-I.}\ \bibnamefont {Lu}},
  \bibinfo {author} {\bibfnamefont {L.~M.}\ \bibnamefont {Aycock}}, \bibinfo
  {author} {\bibfnamefont {D.}~\bibnamefont {Genkina}}, \ and\ \bibinfo
  {author} {\bibfnamefont {I.~B.}\ \bibnamefont {Spielman}},\ }\href {\doibase
  10.1126/science.aaa8515} {\bibfield  {journal} {\bibinfo  {journal}
  {Science}\ }\textbf {\bibinfo {volume} {349}},\ \bibinfo {pages} {1514}
  (\bibinfo {year} {2015})}\BibitemShut {NoStop}%
\bibitem [{\citenamefont {Salerno}\ \emph {et~al.}(2019)\citenamefont
  {Salerno}, \citenamefont {Price}, \citenamefont {Lebrat}, \citenamefont
  {H\"ausler}, \citenamefont {Esslinger}, \citenamefont {Corman}, \citenamefont
  {Brantut},\ and\ \citenamefont {Goldman}}]{PhysRevX.9.041001}%
  \BibitemOpen
  \bibfield  {author} {\bibinfo {author} {\bibfnamefont {G.}~\bibnamefont
  {Salerno}}, \bibinfo {author} {\bibfnamefont {H.~M.}\ \bibnamefont {Price}},
  \bibinfo {author} {\bibfnamefont {M.}~\bibnamefont {Lebrat}}, \bibinfo
  {author} {\bibfnamefont {S.}~\bibnamefont {H\"ausler}}, \bibinfo {author}
  {\bibfnamefont {T.}~\bibnamefont {Esslinger}}, \bibinfo {author}
  {\bibfnamefont {L.}~\bibnamefont {Corman}}, \bibinfo {author} {\bibfnamefont
  {J.-P.}\ \bibnamefont {Brantut}}, \ and\ \bibinfo {author} {\bibfnamefont
  {N.}~\bibnamefont {Goldman}},\ }\href {\doibase 10.1103/PhysRevX.9.041001}
  {\bibfield  {journal} {\bibinfo  {journal} {Phys. Rev. X}\ }\textbf {\bibinfo
  {volume} {9}},\ \bibinfo {pages} {041001} (\bibinfo {year}
  {2019})}\BibitemShut {NoStop}%
\bibitem [{\citenamefont {Ozawa}\ and\ \citenamefont
  {Price}(2019)}]{Ozawa:2019aa}%
  \BibitemOpen
  \bibfield  {author} {\bibinfo {author} {\bibfnamefont {T.}~\bibnamefont
  {Ozawa}}\ and\ \bibinfo {author} {\bibfnamefont {H.~M.}\ \bibnamefont
  {Price}},\ }\href {\doibase 10.1038/s42254-019-0045-3} {\bibfield  {journal}
  {\bibinfo  {journal} {Nature Reviews Physics}\ }\textbf {\bibinfo {volume}
  {1}},\ \bibinfo {pages} {349} (\bibinfo {year} {2019})}\BibitemShut {NoStop}%
\bibitem [{\citenamefont {Price}\ \emph {et~al.}(2020)\citenamefont {Price},
  \citenamefont {Ozawa},\ and\ \citenamefont {Schomerus}}]{HPrice}%
  \BibitemOpen
  \bibfield  {author} {\bibinfo {author} {\bibfnamefont {H.}~\bibnamefont
  {Price}}, \bibinfo {author} {\bibfnamefont {T.}~\bibnamefont {Ozawa}}, \ and\
  \bibinfo {author} {\bibfnamefont {H.}~\bibnamefont {Schomerus}},\ }\href@noop
  {} {\bibfield  {journal} {\bibinfo  {journal} {arXiv:1907.04231}\ } (\bibinfo
  {year} {2020})}\BibitemShut {NoStop}%
\bibitem [{\citenamefont {Price}\ \emph {et~al.}(2017)\citenamefont {Price},
  \citenamefont {Ozawa},\ and\ \citenamefont {Goldman}}]{PhysRevA.95.023607}%
  \BibitemOpen
  \bibfield  {author} {\bibinfo {author} {\bibfnamefont {H.~M.}\ \bibnamefont
  {Price}}, \bibinfo {author} {\bibfnamefont {T.}~\bibnamefont {Ozawa}}, \ and\
  \bibinfo {author} {\bibfnamefont {N.}~\bibnamefont {Goldman}},\ }\href
  {\doibase 10.1103/PhysRevA.95.023607} {\bibfield  {journal} {\bibinfo
  {journal} {Phys. Rev. A}\ }\textbf {\bibinfo {volume} {95}},\ \bibinfo
  {pages} {023607} (\bibinfo {year} {2017})}\BibitemShut {NoStop}%
\bibitem [{\citenamefont {Buser}\ \emph {et~al.}(2020)\citenamefont {Buser},
  \citenamefont {Hubig}, \citenamefont {Schollw{\"o}ck}, \citenamefont
  {Tarruell},\ and\ \citenamefont {Heidrich-Meisner}}]{Cr_V}%
  \BibitemOpen
  \bibfield  {author} {\bibinfo {author} {\bibfnamefont {M.}~\bibnamefont
  {Buser}}, \bibinfo {author} {\bibfnamefont {C.}~\bibnamefont {Hubig}},
  \bibinfo {author} {\bibfnamefont {U.}~\bibnamefont {Schollw{\"o}ck}},
  \bibinfo {author} {\bibfnamefont {L.}~\bibnamefont {Tarruell}}, \ and\
  \bibinfo {author} {\bibfnamefont {F.}~\bibnamefont {Heidrich-Meisner}},\
  }\href@noop {} {\bibfield  {journal} {\bibinfo  {journal} {arXiv:2006.13862}\
  } (\bibinfo {year} {2020})}\BibitemShut {NoStop}%
\bibitem [{\citenamefont {Bravyi}\ \emph {et~al.}(2011)\citenamefont {Bravyi},
  \citenamefont {DiVincenzo},\ and\ \citenamefont {Loss}}]{BRAVYI20112793}%
  \BibitemOpen
  \bibfield  {author} {\bibinfo {author} {\bibfnamefont {S.}~\bibnamefont
  {Bravyi}}, \bibinfo {author} {\bibfnamefont {D.~P.}\ \bibnamefont
  {DiVincenzo}}, \ and\ \bibinfo {author} {\bibfnamefont {D.}~\bibnamefont
  {Loss}},\ }\href {\doibase https://doi.org/10.1016/j.aop.2011.06.004}
  {\bibfield  {journal} {\bibinfo  {journal} {Annals of Physics}\ }\textbf
  {\bibinfo {volume} {326}},\ \bibinfo {pages} {2793 } (\bibinfo {year}
  {2011})}\BibitemShut {NoStop}%
\bibitem [{\citenamefont {Zauner-Stauber}\ \emph {et~al.}(2018)\citenamefont
  {Zauner-Stauber}, \citenamefont {Vanderstraeten}, \citenamefont {Fishman},
  \citenamefont {Verstraete},\ and\ \citenamefont
  {Haegeman}}]{PhysRevB.97.045145}%
  \BibitemOpen
  \bibfield  {author} {\bibinfo {author} {\bibfnamefont {V.}~\bibnamefont
  {Zauner-Stauber}}, \bibinfo {author} {\bibfnamefont {L.}~\bibnamefont
  {Vanderstraeten}}, \bibinfo {author} {\bibfnamefont {M.~T.}\ \bibnamefont
  {Fishman}}, \bibinfo {author} {\bibfnamefont {F.}~\bibnamefont {Verstraete}},
  \ and\ \bibinfo {author} {\bibfnamefont {J.}~\bibnamefont {Haegeman}},\
  }\href {\doibase 10.1103/PhysRevB.97.045145} {\bibfield  {journal} {\bibinfo
  {journal} {Phys. Rev. B}\ }\textbf {\bibinfo {volume} {97}},\ \bibinfo
  {pages} {045145} (\bibinfo {year} {2018})}\BibitemShut {NoStop}%
\bibitem [{\citenamefont {Rossini}\ and\ \citenamefont
  {Fazio}(2012)}]{Rossini_2012}%
  \BibitemOpen
  \bibfield  {author} {\bibinfo {author} {\bibfnamefont {D.}~\bibnamefont
  {Rossini}}\ and\ \bibinfo {author} {\bibfnamefont {R.}~\bibnamefont
  {Fazio}},\ }\href {\doibase 10.1088/1367-2630/14/6/065012} {\bibfield
  {journal} {\bibinfo  {journal} {New Journal of Physics}\ }\textbf {\bibinfo
  {volume} {14}},\ \bibinfo {pages} {065012} (\bibinfo {year}
  {2012})}\BibitemShut {NoStop}%
\bibitem [{\citenamefont {K\"uhner}\ and\ \citenamefont
  {Monien}(1998)}]{PhysRevB.58.R14741}%
  \BibitemOpen
  \bibfield  {author} {\bibinfo {author} {\bibfnamefont {T.~D.}\ \bibnamefont
  {K\"uhner}}\ and\ \bibinfo {author} {\bibfnamefont {H.}~\bibnamefont
  {Monien}},\ }\href {\doibase 10.1103/PhysRevB.58.R14741} {\bibfield
  {journal} {\bibinfo  {journal} {Phys. Rev. B}\ }\textbf {\bibinfo {volume}
  {58}},\ \bibinfo {pages} {R14741} (\bibinfo {year} {1998})}\BibitemShut
  {NoStop}%
\bibitem [{\citenamefont {Buyskikh}\ \emph {et~al.}(2019)\citenamefont
  {Buyskikh}, \citenamefont {Tagliacozzo}, \citenamefont {Schuricht},
  \citenamefont {Hooley}, \citenamefont {Pekker},\ and\ \citenamefont
  {Daley}}]{PhysRevA.100.023627}%
  \BibitemOpen
  \bibfield  {author} {\bibinfo {author} {\bibfnamefont {A.~S.}\ \bibnamefont
  {Buyskikh}}, \bibinfo {author} {\bibfnamefont {L.}~\bibnamefont
  {Tagliacozzo}}, \bibinfo {author} {\bibfnamefont {D.}~\bibnamefont
  {Schuricht}}, \bibinfo {author} {\bibfnamefont {C.~A.}\ \bibnamefont
  {Hooley}}, \bibinfo {author} {\bibfnamefont {D.}~\bibnamefont {Pekker}}, \
  and\ \bibinfo {author} {\bibfnamefont {A.~J.}\ \bibnamefont {Daley}},\ }\href
  {\doibase 10.1103/PhysRevA.100.023627} {\bibfield  {journal} {\bibinfo
  {journal} {Phys. Rev. A}\ }\textbf {\bibinfo {volume} {100}},\ \bibinfo
  {pages} {023627} (\bibinfo {year} {2019})}\BibitemShut {NoStop}%
\bibitem [{\citenamefont {Verstraete}\ \emph {et~al.}(2004)\citenamefont
  {Verstraete}, \citenamefont {Garc\'{\i}a-Ripoll},\ and\ \citenamefont
  {Cirac}}]{PhysRevLett.93.207204}%
  \BibitemOpen
  \bibfield  {author} {\bibinfo {author} {\bibfnamefont {F.}~\bibnamefont
  {Verstraete}}, \bibinfo {author} {\bibfnamefont {J.~J.}\ \bibnamefont
  {Garc\'{\i}a-Ripoll}}, \ and\ \bibinfo {author} {\bibfnamefont {J.~I.}\
  \bibnamefont {Cirac}},\ }\href {\doibase 10.1103/PhysRevLett.93.207204}
  {\bibfield  {journal} {\bibinfo  {journal} {Phys. Rev. Lett.}\ }\textbf
  {\bibinfo {volume} {93}},\ \bibinfo {pages} {207204} (\bibinfo {year}
  {2004})}\BibitemShut {NoStop}%
\bibitem [{\citenamefont {Phien}\ \emph {et~al.}(2013)\citenamefont {Phien},
  \citenamefont {Vidal},\ and\ \citenamefont {McCulloch}}]{PhysRevB.88.035103}%
  \BibitemOpen
  \bibfield  {author} {\bibinfo {author} {\bibfnamefont {H.~N.}\ \bibnamefont
  {Phien}}, \bibinfo {author} {\bibfnamefont {G.}~\bibnamefont {Vidal}}, \ and\
  \bibinfo {author} {\bibfnamefont {I.~P.}\ \bibnamefont {McCulloch}},\ }\href
  {\doibase 10.1103/PhysRevB.88.035103} {\bibfield  {journal} {\bibinfo
  {journal} {Phys. Rev. B}\ }\textbf {\bibinfo {volume} {88}},\ \bibinfo
  {pages} {035103} (\bibinfo {year} {2013})}\BibitemShut {NoStop}%
\bibitem [{\citenamefont {Phien}\ \emph {et~al.}(2012)\citenamefont {Phien},
  \citenamefont {Vidal},\ and\ \citenamefont {McCulloch}}]{PhysRevB.86.245107}%
  \BibitemOpen
  \bibfield  {author} {\bibinfo {author} {\bibfnamefont {H.~N.}\ \bibnamefont
  {Phien}}, \bibinfo {author} {\bibfnamefont {G.}~\bibnamefont {Vidal}}, \ and\
  \bibinfo {author} {\bibfnamefont {I.~P.}\ \bibnamefont {McCulloch}},\ }\href
  {\doibase 10.1103/PhysRevB.86.245107} {\bibfield  {journal} {\bibinfo
  {journal} {Phys. Rev. B}\ }\textbf {\bibinfo {volume} {86}},\ \bibinfo
  {pages} {245107} (\bibinfo {year} {2012})}\BibitemShut {NoStop}%
\bibitem [{\citenamefont {Landau}(1947)}]{LLSS}%
  \BibitemOpen
  \bibfield  {author} {\bibinfo {author} {\bibfnamefont {L.}~\bibnamefont
  {Landau}},\ }\href@noop {} {\bibfield  {journal} {\bibinfo  {journal} {J.
  Phys. U.S.S.R.}\ }\textbf {\bibinfo {volume} {11}},\ \bibinfo {pages} {91}
  (\bibinfo {year} {1947})}\BibitemShut {NoStop}%
\bibitem [{\citenamefont {Feynman}(1954)}]{PhysRev.94.262}%
  \BibitemOpen
  \bibfield  {author} {\bibinfo {author} {\bibfnamefont {R.~P.}\ \bibnamefont
  {Feynman}},\ }\href {\doibase 10.1103/PhysRev.94.262} {\bibfield  {journal}
  {\bibinfo  {journal} {Phys. Rev.}\ }\textbf {\bibinfo {volume} {94}},\
  \bibinfo {pages} {262} (\bibinfo {year} {1954})}\BibitemShut {NoStop}%
\bibitem [{\citenamefont {Chomaz}\ \emph {et~al.}(2018)\citenamefont {Chomaz},
  \citenamefont {van Bijnen}, \citenamefont {Petter}, \citenamefont {Faraoni},
  \citenamefont {Baier}, \citenamefont {Becher}, \citenamefont {Mark},
  \citenamefont {W{\"a}chtler}, \citenamefont {Santos},\ and\ \citenamefont
  {Ferlaino}}]{Chomaz:2018aa}%
  \BibitemOpen
  \bibfield  {author} {\bibinfo {author} {\bibfnamefont {L.}~\bibnamefont
  {Chomaz}}, \bibinfo {author} {\bibfnamefont {R.~M.~W.}\ \bibnamefont {van
  Bijnen}}, \bibinfo {author} {\bibfnamefont {D.}~\bibnamefont {Petter}},
  \bibinfo {author} {\bibfnamefont {G.}~\bibnamefont {Faraoni}}, \bibinfo
  {author} {\bibfnamefont {S.}~\bibnamefont {Baier}}, \bibinfo {author}
  {\bibfnamefont {J.~H.}\ \bibnamefont {Becher}}, \bibinfo {author}
  {\bibfnamefont {M.~J.}\ \bibnamefont {Mark}}, \bibinfo {author}
  {\bibfnamefont {F.}~\bibnamefont {W{\"a}chtler}}, \bibinfo {author}
  {\bibfnamefont {L.}~\bibnamefont {Santos}}, \ and\ \bibinfo {author}
  {\bibfnamefont {F.}~\bibnamefont {Ferlaino}},\ }\href {\doibase
  10.1038/s41567-018-0054-7} {\bibfield  {journal} {\bibinfo  {journal} {Nature
  Physics}\ }\textbf {\bibinfo {volume} {14}},\ \bibinfo {pages} {442}
  (\bibinfo {year} {2018})}\BibitemShut {NoStop}%
\bibitem [{\citenamefont {Daley}\ and\ \citenamefont
  {Simon}(2014)}]{PhysRevA.89.053619}%
  \BibitemOpen
  \bibfield  {author} {\bibinfo {author} {\bibfnamefont {A.~J.}\ \bibnamefont
  {Daley}}\ and\ \bibinfo {author} {\bibfnamefont {J.}~\bibnamefont {Simon}},\
  }\href {\doibase 10.1103/PhysRevA.89.053619} {\bibfield  {journal} {\bibinfo
  {journal} {Phys. Rev. A}\ }\textbf {\bibinfo {volume} {89}},\ \bibinfo
  {pages} {053619} (\bibinfo {year} {2014})}\BibitemShut {NoStop}%
\bibitem [{\citenamefont {Giamarchi}(2003)}]{Gi_Lutt}%
  \BibitemOpen
  \bibfield  {author} {\bibinfo {author} {\bibfnamefont {T.}~\bibnamefont
  {Giamarchi}},\ }\href@noop {} {\emph {\bibinfo {title} {Quantum Physics in
  One Dimension}}},\ International Series of Monographs on Physics\ (\bibinfo
  {publisher} {Oxford University Press},\ \bibinfo {year} {2003})\BibitemShut
  {NoStop}%
\bibitem [{\citenamefont {Kane}\ and\ \citenamefont
  {Fisher}(1992)}]{PhysRevLett.68.1220}%
  \BibitemOpen
  \bibfield  {author} {\bibinfo {author} {\bibfnamefont {C.~L.}\ \bibnamefont
  {Kane}}\ and\ \bibinfo {author} {\bibfnamefont {M.~P.~A.}\ \bibnamefont
  {Fisher}},\ }\href {\doibase 10.1103/PhysRevLett.68.1220} {\bibfield
  {journal} {\bibinfo  {journal} {Phys. Rev. Lett.}\ }\textbf {\bibinfo
  {volume} {68}},\ \bibinfo {pages} {1220} (\bibinfo {year}
  {1992})}\BibitemShut {NoStop}%
\bibitem [{\citenamefont {Kane}\ and\ \citenamefont
  {Fisher}(1996)}]{PhysRevLett.76.3192}%
  \BibitemOpen
  \bibfield  {author} {\bibinfo {author} {\bibfnamefont {C.~L.}\ \bibnamefont
  {Kane}}\ and\ \bibinfo {author} {\bibfnamefont {M.~P.~A.}\ \bibnamefont
  {Fisher}},\ }\href {\doibase 10.1103/PhysRevLett.76.3192} {\bibfield
  {journal} {\bibinfo  {journal} {Phys. Rev. Lett.}\ }\textbf {\bibinfo
  {volume} {76}},\ \bibinfo {pages} {3192} (\bibinfo {year}
  {1996})}\BibitemShut {NoStop}%
\bibitem [{\citenamefont {Imambekov}\ \emph {et~al.}(2012)\citenamefont
  {Imambekov}, \citenamefont {Schmidt},\ and\ \citenamefont
  {Glazman}}]{RevModPhys.84.1253}%
  \BibitemOpen
  \bibfield  {author} {\bibinfo {author} {\bibfnamefont {A.}~\bibnamefont
  {Imambekov}}, \bibinfo {author} {\bibfnamefont {T.~L.}\ \bibnamefont
  {Schmidt}}, \ and\ \bibinfo {author} {\bibfnamefont {L.~I.}\ \bibnamefont
  {Glazman}},\ }\href {\doibase 10.1103/RevModPhys.84.1253} {\bibfield
  {journal} {\bibinfo  {journal} {Rev. Mod. Phys.}\ }\textbf {\bibinfo {volume}
  {84}},\ \bibinfo {pages} {1253} (\bibinfo {year} {2012})}\BibitemShut
  {NoStop}%
\bibitem [{\citenamefont {Fabbri}\ \emph {et~al.}(2015)\citenamefont {Fabbri},
  \citenamefont {Panfil}, \citenamefont {Cl\'ement}, \citenamefont {Fallani},
  \citenamefont {Inguscio}, \citenamefont {Fort},\ and\ \citenamefont
  {Caux}}]{PhysRevA.91.043617}%
  \BibitemOpen
  \bibfield  {author} {\bibinfo {author} {\bibfnamefont {N.}~\bibnamefont
  {Fabbri}}, \bibinfo {author} {\bibfnamefont {M.}~\bibnamefont {Panfil}},
  \bibinfo {author} {\bibfnamefont {D.}~\bibnamefont {Cl\'ement}}, \bibinfo
  {author} {\bibfnamefont {L.}~\bibnamefont {Fallani}}, \bibinfo {author}
  {\bibfnamefont {M.}~\bibnamefont {Inguscio}}, \bibinfo {author}
  {\bibfnamefont {C.}~\bibnamefont {Fort}}, \ and\ \bibinfo {author}
  {\bibfnamefont {J.-S.}\ \bibnamefont {Caux}},\ }\href {\doibase
  10.1103/PhysRevA.91.043617} {\bibfield  {journal} {\bibinfo  {journal} {Phys.
  Rev. A}\ }\textbf {\bibinfo {volume} {91}},\ \bibinfo {pages} {043617}
  (\bibinfo {year} {2015})}\BibitemShut {NoStop}%
\bibitem [{\citenamefont {Chitra}\ and\ \citenamefont
  {Giamarchi}(1997)}]{PhysRevB.55.5816}%
  \BibitemOpen
  \bibfield  {author} {\bibinfo {author} {\bibfnamefont {R.}~\bibnamefont
  {Chitra}}\ and\ \bibinfo {author} {\bibfnamefont {T.}~\bibnamefont
  {Giamarchi}},\ }\href {\doibase 10.1103/PhysRevB.55.5816} {\bibfield
  {journal} {\bibinfo  {journal} {Phys. Rev. B}\ }\textbf {\bibinfo {volume}
  {55}},\ \bibinfo {pages} {5816} (\bibinfo {year} {1997})}\BibitemShut
  {NoStop}%
\end{thebibliography}%

\end{document}